\documentclass[12pt]{iopart}
\usepackage{iopams} 
 
\usepackage{color}
\usepackage{ulem}
\usepackage{hyperref}
\usepackage{graphicx}
\usepackage{subfig}

\begin{document}

\title{Impurity radiation seeding of neoclassical tearing mode growth}

\author{Shiyong Zeng}
\address{State Key Laboratory of Advanced Electromagnetic Engineering and Technology, International Joint Research Laboratory of Magnetic Confinement Fusion and Plasma Physics, School of Electrical and Electronic Engineering, Huazhong University of Science and Technology, Wuhan, Hubei 430074, China}
\vspace{10pt}

\author{Ping Zhu}
\address{State Key Laboratory of Advanced Electromagnetic Engineering and Technology, International Joint Research Laboratory of Magnetic Confinement Fusion and Plasma Physics, School of Electrical and Electronic Engineering, Huazhong University of Science and Technology, Wuhan, Hubei 430074, China}
\address{Department of Nuclear Engineering and Engineering Physics, University of Wisconsin-Madison, Madison, Wisconsin 53706, USA}
\ead{zhup@hust.edu.cn}
\vspace{10pt}

\author{Eric C. Howell}
\address{Tech-X Corporation, 5621 Arapahoe Avenue Suite A, Boulder, Colorado 80303, USA}

\vspace{10pt}
\begin{indented}
\item[]\today
\end{indented}

\begin{abstract}
The physics of neoclassical tearing mode (NTM) is of great concern to the tokamak plasma stability and performance, especially in the burning plasma regime. Whereas a great deal about the different seeding mechanisms have been understood, and in many situations the seed event can be clearly identified, the potential seeding process of NTM due to the resistive tearing instability driven by the impurity radiation cooling still needs more studies. Recent NIMROD simulations have demonstrated that the local impurity radiation cooling can drive the seed island growth and trigger the subsequent onset of neoclassical tearing mode instability. The seed island is mainly driven by the local helical perturbation of the diamagnetic current induced by the perturbed pressure gradient as a result of the impurity radiative cooling on the rational surface. A heuristic closure for the neoclassical viscosity is adopted, and the seed island is further driven by the perturbed bootstrap current induced from the neoclassical electron viscous stress in the extended Ohm’s law. The growth rate of the NTM in simulations is found proportional to the electron neoclassical viscosity, and a theoretical neoclassical driving term is adopted to account for the nonlinear neoclassical island growth in the simulations. 
\end{abstract}

%
%
%
%
%

\section{Introduction}
\label{Sec:introduction}
The physics of neoclassical tearing mode (NTM) is of great concern to the tokamak plasma stability and performance, especially in the burning plasma regime, since it was first observed on TFTR \cite{Chang1995}. It is well known that the NTMs are detrimental to the fusion plasmas \cite{Sauter1997,Hegna1998,Buttery_2000,Haye_2022}, which degrade the confinement of plasma by short-circuit the nested flux surfaces in presence of the magnetic island formation \cite{Meskat_2001,Oikawa_2005_PhysRevLett.94.125003}, and even lead to uncontrolled discharge termination, i.e. the plasma disruption. Thus NTMs can bring down the highest achievable value of $\beta_N=\beta(\%)/[I_p(MA)/a(m)B_0(T)]$ much below the ideal limit \cite{Gates1997,Sauter_2002,Maraschek_2003}, i.e. $\beta_N\sim4l_i$, where $\beta=2\mu_0\left\langle p\right\rangle/B_0^2$ is the ratio of plasma thermal pressure to the magnetic pressure, $\left\langle p\right\rangle$ the volume averaged plasma thermal pressure, $a$ the minor radius, $I_p$ the plasma current, $B_0$ the magnetic field and $l_i$ the plasma internal inductance. 

On the one hand, the empirical scaling shows the marginal $\beta$ limit for the NTM growth is almost linearly proportional to the local poloidal ion gyro-radius $\rho_{pi}^{*}$
\cite{Maraschek_2003,Buttery_2003_onset}, which predicts very low $\beta$ threshold on future large tokamak devices, e.g. $\rho_{pi}^{*}\sim 1\times 10^{-3}$ in ITER \cite{Haye_2000}. In addition, the seed island width required for NTM onset is predicted to be $1\sim2$cm in ITER plasma \cite{Haye_2022}. On the other hand, the creation mechanism of seed island required for the NTM onset and growth remains open to more work, such as the development of theory based predictive capabilities and the explanation of why certain seed events trigger an NTM while others do not.
Previous studies have proposed the seed island triggers by precursor MHD instabilities, such as the sawtooth \cite{Gude_1999,Sauter_2022_PhysRevLett.88.105001,Canal_2013},
the fishbone \cite{Gude_1999},
the edge localized mode \cite{Haye_2022},
and the infernal mode \cite{Kleiner_2016}.
The spontaneous NTM onset in absence of detectable precursor MHD events has been discussed as well in \cite{Fredrickson_2002,Brennan2002,Brennan2003,Brennan_2005}, where it is reported that the tearing index $\Delta'$ becomes extreme large and positive before the NTM onset as the equilibrium approaches the ideal stability boundary.
TCV experiments demonstrate that the NTM can grow from the current driven tearing mode \cite{Reimerdes2002}.
In addition, the seeding mechanisms by mode coupling \cite{Bardoczi_PhysRevLett.127.055002,Nave_2003},
resonant magnetic perturbations \cite{Buttery_2001,Qu_2022,Yu_2012},
and turbulence \cite{Muraglia_PhysRevLett.107.095003,Muraglia_2017,Muraglia_2021}
are also suggested to be responsible for the NTM onset.
Two main theory models have been developed to account for the onset of NTM growth, namely the finite thermal transport model and the polarization current model
\cite{Zabiego_1997,Haye_1998,Buttery_2003}.
Nevertheless, detecting the seed island is very difficult in experiments, and the driving terms in NTM model need to be exactly measured on the local rational surface, both of which contribute to the challenge in comparing between theory model and experimental data.

Meanwhile, the impurity radiation is of great importance to the steady state operation of tokamak plasmas, and plays key role in driving tearing mode instability, which has been confirmed in a number of experiments \cite{Suttrop_1997,Salzedas2002,Delgado-Aparicio_2011,Xu_2017}. A nonlinear theory model based on cylindrical geometry is developed to interpret this thermo-resistive tearing instability \cite{Gates2012,White2015}, and the impurity radiation driving tearing mode growth has been studied numerically using NIMROD \cite{Zeng_2023_tm,Zeng_2023,Zeng_2023_PPCF}, JOREK \cite{Nardon_2017,Wieschollek_2022}, M3D-C1 \cite{Teng_2018}, MDC \cite{JIANG_2022,Jiang_2023} and CLT \cite{Xu_CLT_2020}.
Naturally, the impurity radiation induced resistive tearing mode could be a potential candidate for the seed island of NTM growth. For example, the JET experiments exhibit the onset of tearing mode as a consequence of the current density profile modification following the changes in the electron temperature profile due to radiative cooling \cite{Pucella_2021}. However, the experimental observations only indicate the correlation between the impurity radiation and the potential NTM growth, whereas the detailed underlying physics maybe less clear.


This work intends to explore an radiation-driven seeding mechanism for the NTM onset, by demonstrating the local impurity radiation cooling seeded island and the subsequent bootstrap current driven neoclassical tearing growth. Section \ref{Sec:simulation model and equilibrium} introduces the simulation model and setup, as well as the equilibrium adopted in the simulations. Section \ref{Sec:impurity radiation induced seed} reports the local impurity radiation cooling induced seed island growth. Section \ref{Sec:impurity radiation seeded NTM} focuses on the subsequent island growth driven by the perturbed bootstrap current. The simulation results are analyzed in comparison with theory models in both Section \ref{Sec:impurity radiation induced seed} and \ref{Sec:impurity radiation seeded NTM}. Finally, conclusions are provided in Section \ref{Sec:conclusion}.

\section{Simulation model and equilibrium}
\label{Sec:simulation model and equilibrium}

\subsection{Simulation model}
Our simulations are based on the single-fluid resistive MHD model implemented in the NIMROD code \cite{Sovinec2004}, which incorporates an atomic physics module ported from the KPRAD code to calculate the particle and energy sources due to the impurity ionization, recombination and radiation processes \cite{KPRAD,Izzo_2013},
and a heuristic closure for the neoclassical viscosities \cite{Gianakon_2002,Howell_2022}
\begin{eqnarray}
	\rho \frac{{\rm d}\bi{V}}{{\rm d}t} = - \nabla p + \bi{J}\times\bi{B} + \nabla\cdot\left(\rho\nu\nabla\bi{V}\right) - \nabla\cdot\bi{\Pi}_i,
	\label{eq:momentum}
	\\
	\frac{{\rm d} n_i}{{\rm d}t} + n_i \nabla \cdot \bi{V} = \nabla \cdot (D \nabla n_i) + S_{ion/rec},
	\label{eq:contiune2}
	\\
	\frac{{\rm d} n_{Z}}{{\rm d}t} + n_Z \nabla \cdot \bi{V} = \nabla \cdot (D \nabla n_Z) + S_{ion/rec},
	\label{eq:contiune3}
	\\
	n_e \frac{{\rm d}T_e}{{\rm d}t} = (\gamma - 1)[n_e T_e \nabla \cdot \bi{V} + \nabla \cdot \bi{q_e} - Q],
	\label{eq:temperature}
	\\
	\bi{q}_e = -n_e[\kappa_{\parallel} \bi{b}\bi{b} + \kappa_{\perp} (\mathcal{I} - \bi{b}\bi{b})] \cdot \nabla T_e,
	\label{eq:heat_flux}
	\\
	\frac{\partial \bi{B}}{\partial t} = -\nabla\times\bi{E},
	\label{eq:Faraday law}
	\\
	\bi{E} = -\bi{V}\times\bi{B} + \eta\bi{J} - \frac{1}{en_e}\nabla\cdot\bi{\Pi}_e,
	\label{eq:Ohm law}
	\\
	\nabla\times\bi{B} = \mu_0\bi{J}.
	\label{eq:Ampere law}
\end{eqnarray}	
Here, $n_i$, $n_e$, and $n_Z$ are the main ion, electron, and impurity ion number densities respectively, $e$, $\rho$, $\bi{V}$, $\bi{J}$, and $p$ the electron charge, the plasma mass density, velocity, current density, and pressure respectively, $T_e$ and $\bi{q}_e$ the electron temperature and heat flux respectively, $D$, $\nu$, $\eta$, and $\kappa_{\parallel}$ ($\kappa_{\perp}$) the plasma diffusivity, kinematic viscosity, resistivity, and parallel (perpendicular) thermal conductivity respectively, $\gamma = 5/3$ the adiabatic index, $S_{ion/rec}$ the density source, $Q$ the energy source, $\bi{E}$ ($\bi{B}$) the electric (magnetic) field, $\bi{b}=\bi{B}/B$, $\mathcal{I}$ the unit dyadic tensor,
and $\bi{\Pi}_e$ ($\bi{\Pi}_i$) the electron (ion) neoclassical viscous stress tensor.

\begin{figure}[ht]
	\begin{center}
		\includegraphics[width=0.35\linewidth]{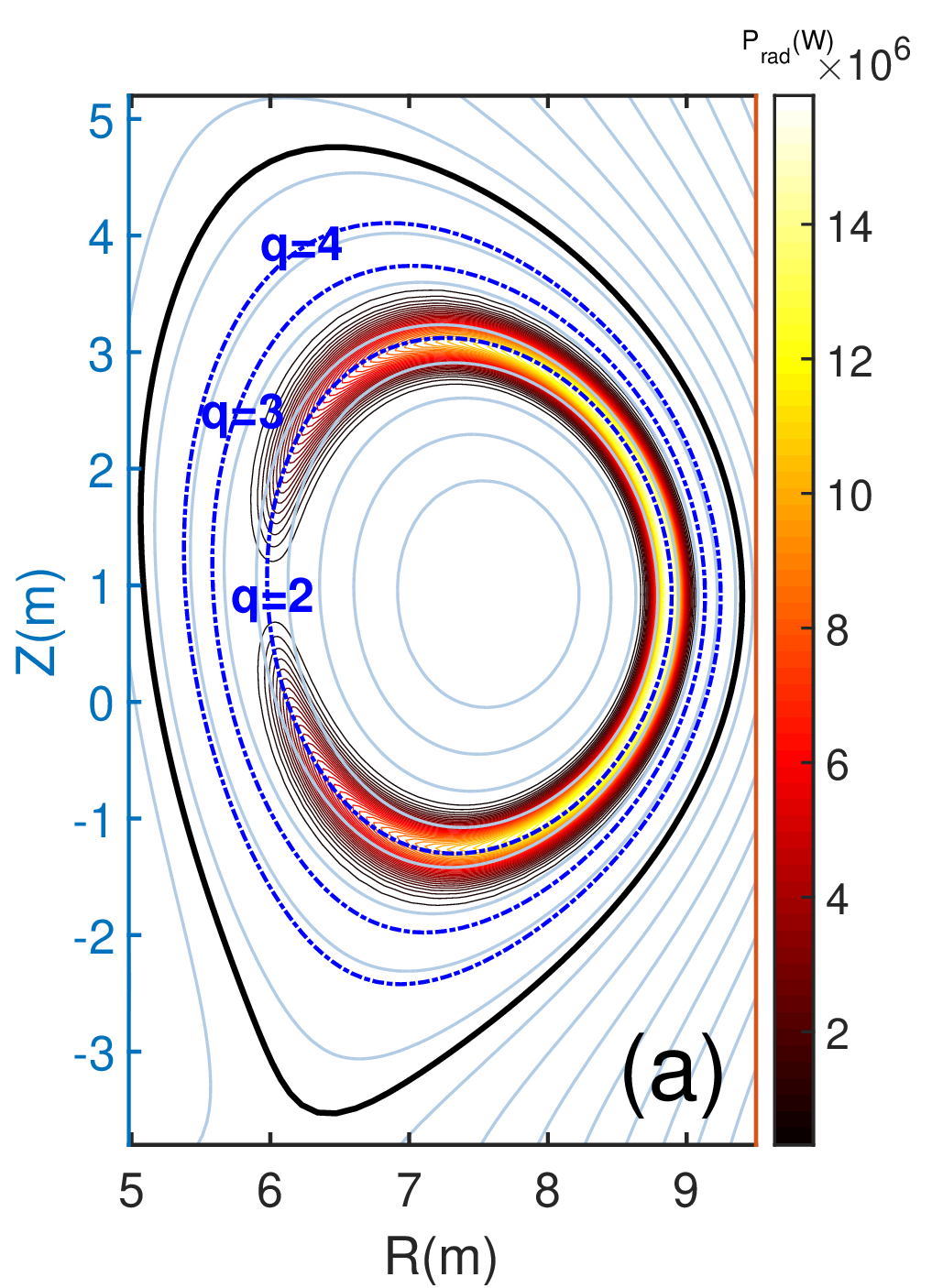}
		\includegraphics[width=0.55\linewidth]{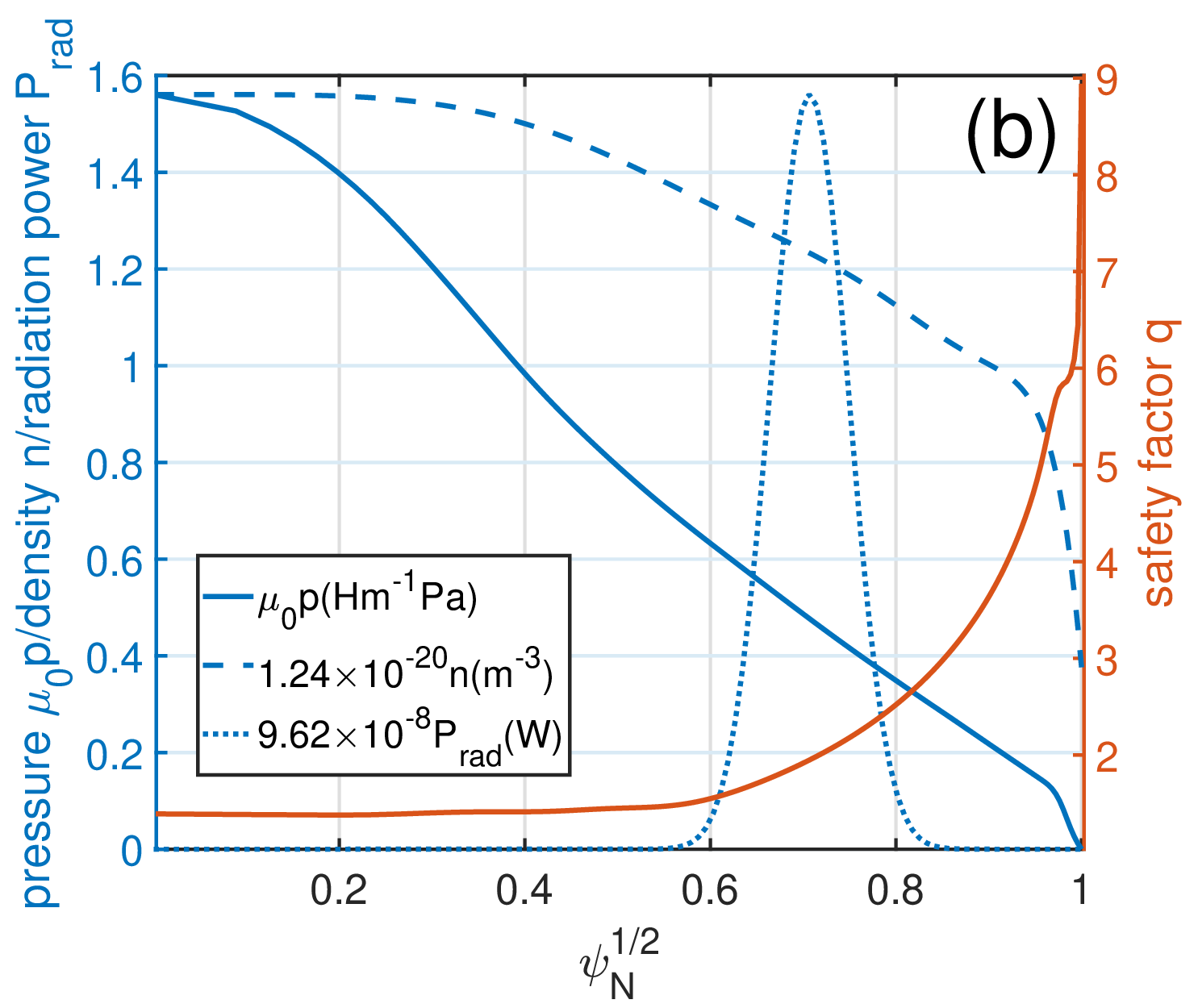}
	\end{center}
	\caption{(a) Contours of the initial equilibrium flux $\psi$ (grey solid lines) and impurity radiation power distribution (flushed color, color bar in unit $W$) in the poloidal plane, where the equilibrium $q=4,3,2$ surfaces are denoted as blue dashed lines and the boundary of simulation domain is denoted as black solid line. (b) The equilibrium profiles of plasma pressure $\mu_0p$ (blue solid line), number density $n$ (blue dashed line), the impurity radiation power (blue dotted line), and safety factor $q$ (orange solid line)  as functions of $\sqrt{\psi_N}$, where $\psi_N$ is the normalized equilibrium poloidal flux.}
	\label{fig: eq profiles}
\end{figure}

The KPRAD module updates the impurity density of different charge state $Z$ at every time step, and the source terms $S_{ion/rec}$ in Eqs. (\ref{eq:contiune2}) and (\ref{eq:contiune3}) come from ionization and recombination.
The energy source term $Q$ in Eq. (\ref{eq:temperature}) consists of the impurity power loss calculated in the KPRAD module including those from line radiation ($\propto f_{line}(T_e) n_en_Z$), ionization ($\propto f_{ion}(T_e)n_e n_Z$), recombination ($\propto f_{rec}(T_e)n_en_ZT_e$), and bremsstrahlung ($\propto n_e^2T_e^{1/2} Z_{eff}$, $Z_{eff}$ is the effective charge number), as well as the Ohmic heating power $\eta J^2$.
The plasma pressure profile can be largely modified through the temperature and density perturbations in the presence of impurity radiation cooling, as well as the Spitzer resistivity ($\eta \propto T_e^{-3/2}$) adopted in the extended Ohm's law (\ref{eq:Ohm law}).
Both the plasma ion, electron and impurity species share the same temperature and velocity, which is based on the assumptions of instantaneous thermal equilibration and sufficient collisions due to radiative cooling.
Quasi-neutralization $n_e=n_i+\sum Zn_Z$ is also assumed.
The impurity source at the beginning of simulations ($t=0$ms) is the neutral Neon gas and is localized on the $q=2$ rational surface (Fig. \ref{fig: eq profiles}a), which emulates the effects of the pellet deposition itself inside the plasma only, while neglecting those of the associated penetration and ablation processes.

The heuristic closures for the neoclassical ion and electron stress tensors in Eq. (\ref{eq:momentum}) and Ohm's law (\ref{eq:Ohm law}) are
\begin{eqnarray}
	\nabla\cdot\bi{\Pi}_i = \mu_i n_i m_i \left\langle B_{eq}^2\right\rangle \frac{\left(\bi{V}-\bi{V}_{eq} \right)\cdot\bi{e}_{\theta}}{\left( \bi{B}_{eq}\cdot\bi{e}_{\theta}\right)^2}\bi{e}_{\theta},
	\label{eq:neoclassical ion stress}
	\\
	\nabla\cdot\bi{\Pi}_e = -\mu_e \frac{m_e}{e} \left\langle B_{eq}^2\right\rangle \frac{\left(\bi{J}-\bi{J}_{eq} \right)\cdot\bi{e}_{\theta}}{\left( \bi{B}_{eq}\cdot\bi{e}_{\theta}\right)^2}\bi{e}_{\theta},
	\label{eq:neoclassical elec stress}
\end{eqnarray}
respectively. Here the coefficient $\mu_i$ ($\mu_e$) is the neoclassical ion (electron) poloidal flow damping rate and is set to be a constant in the current model, the bracket $\left\langle...\right\rangle$ represents the flux surface averaged value and the subscript ``eq'' denotes equilibrium quantities, $\bi{e}_{\theta}$ is the flux-aligned poloidal basis vector.
Eqs. (\ref{eq:neoclassical ion stress}) and (\ref{eq:neoclassical elec stress}) are simplifications of the viscous stress tensor for a generic particle species $\alpha$ \cite{Gianakon_2002}
\begin{equation}
	\nabla\cdot\bi{\Pi}_{\alpha} = \mu_{\alpha} n_{\alpha}m_{\alpha} \left\langle B_{eq}^2\right\rangle \frac{\bi{V}_{\alpha}\cdot\bi{e}_{\theta}}{\left(\bi{B}_{eq}\cdot\bi{e}_{\theta}\right)^2}\bi{e}_{\theta},
	\label{eq:neoclassical stress tensor}
\end{equation}
where $\bi{V}_{\alpha}$ is the fluid velocity of the species $\alpha$ in the moving frame of reference in presence of an equilibrium flow velocity $\bi{V}_{eq}$.
This form is motivated by the observation that the dominant neoclassical effect is a drag between the trapped and passing particles which results in a damping force along the poloidal direction in an axisymmetric device. The ion flow velocity is essentially the single fluid flow velocity, i.e. $\bi{V}_i=\bi{V}-\bi{V}_{eq}$ in the ion stress tensor, ignoring effects of the tiny electron-ion mass ratio, i.e. $m_e/m_i\ll1$. The assumption of electron flow velocity $\bi{V}_e=-\frac{1}{en_e}\bi{J}$ is adopted in the electron stress tensor in the limit of strong ion poloidal flow damping, i.e. $\bi{V}_i\approx0$ due to $\mu_i\gg1$.
The heuristic ion force in Eq. (\ref{eq:momentum}) causes the damping of poloidal flow and gives rise to stabilization effect on the tearing mode growth. The heuristic electron force in the extended Ohm's law (\ref{eq:Ohm law}) is responsible for the perturbed bootstrap current $\delta \bi{J}_{\parallel}(\psi)=\frac{1}{\eta en_e}\left\langle\nabla\cdot\bi{\Pi}_e\right\rangle_{\parallel}$ that drives the neoclassical tearing mode growth.
The heuristic closure is able to capture the main characteristics of NTM growth \cite{Gianakon_2002,Howell_2022}, and thus may serve as a proxy to the more complete kinetic-MHD models that are still under development \cite{Jepson_2021}.

\subsection{CFETR equilibrium and simulation setup}
The simulations are based on the hybrid scenario design of China Fusion Engineering Test Reactor (CFETR) \cite{Zhuang_2019}, where the equilibrium configuration as well as the impurity radiation power distribution resulting from the impurity deposition on the local rational surface is shown in Figs. \ref{fig: eq profiles}. Other key parameters of the equilibrium can be found in Table \ref{tb:equilibrium parameters}.
\begin{table}
	\caption{\label{tb:equilibrium parameters}Key parameters of the equilibrium}
	\footnotesize\rm
	\begin{tabular*}{\textwidth}{@{}l*{15}{@{\extracolsep{0pt plus12pt}}l}}
		\br
		Parameter & Symbol & Value & Unit\\
		\mr
		Minor radius & $a$ & $2.205$ & m\\
		Major radius & $R_0$ & $7.228$ & m\\
		Plasma current & $I_p$ & $1.267\times10^1$ & MA\\
		Toroidal magnetic field & $B_t$ & $6.503$ & T\\
		Central safety factor & $q_0$ & $1.380$ & Dimensionless\\
		Edge safety factor & $q_{95}$ & $6.039$ & Dimensionless\\
		Central electron temperature & $T_{e,core}$ & $3.082\times10^1$ & keV\\
		Central electron density & $n_{e,core}$ & $1.258\times10^{20}$ & m$^{-3}$\\
		Edge electron temperature & $T_{e,edge}$ & $2.862$ & keV\\
		\br
	\end{tabular*}
\end{table}
The tokamak plasma region in the simulation domain is surrounded by a perfectly conducting wall without a vacuum region. The plasma framework is static without flow $V_{eq}=0$m/s. A constant diffusivity $D=1$ m$^2$s$^{-1}$ is adopted for all particle species, and anisotropic and constant thermal conductivities $\kappa_{\parallel}=1\times10^{10}$ m$^2$s$^{-1}$, $\kappa_{\perp}=1$ m$^2$s$^{-1}$ are used. The kinematic viscosity $\nu=1\times10^3$ m$^2$s$^{-1}$ is used for numerical stability. 
The simulations use $42\times48$ finite elements with the fifth order Lagrange polynomial basis functions in the poloidal plane, and two Fourier modes with toroidal mode numbers $n=0-1$ are considered in the toroidal direction.

\begin{figure}[ht]
	\begin{center}
		\includegraphics[width=0.75\linewidth]{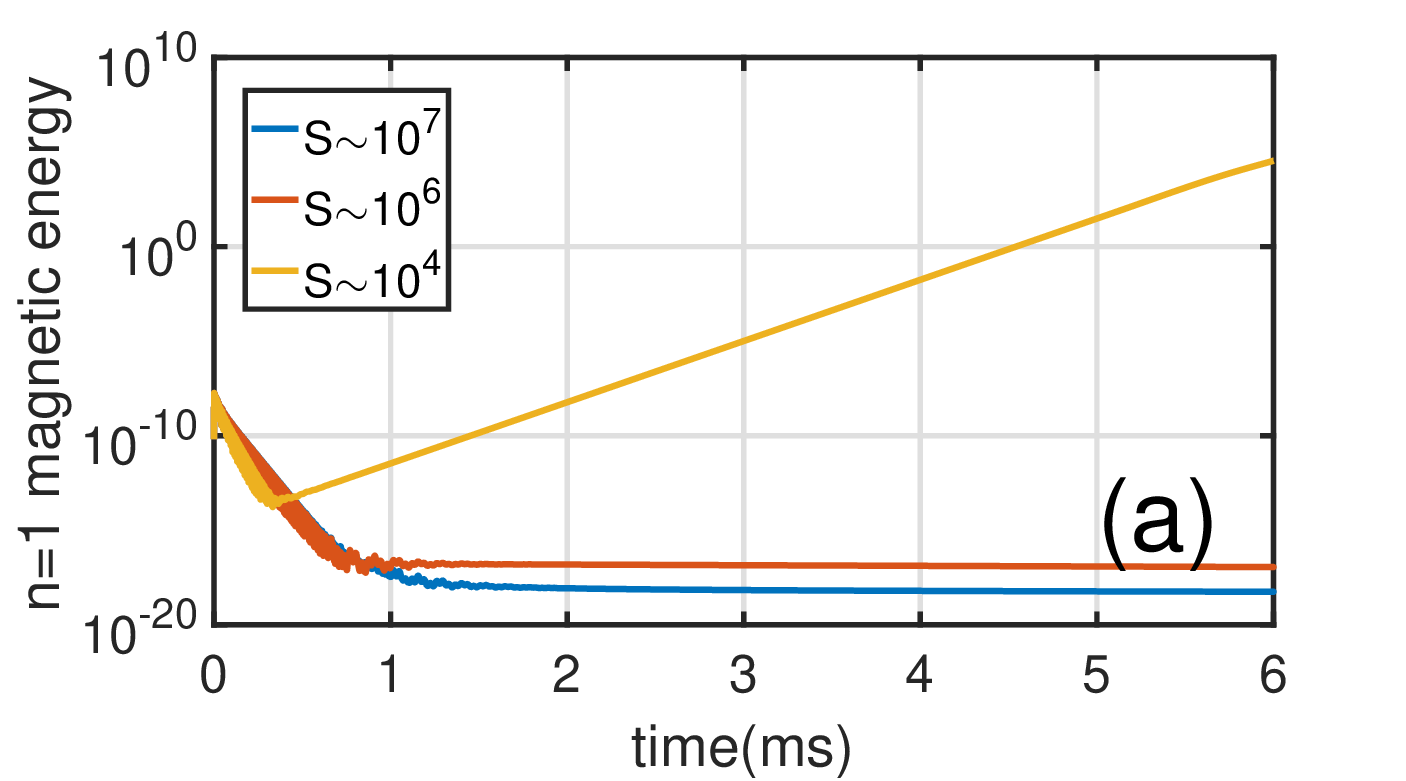}
		\includegraphics[width=0.75\linewidth]{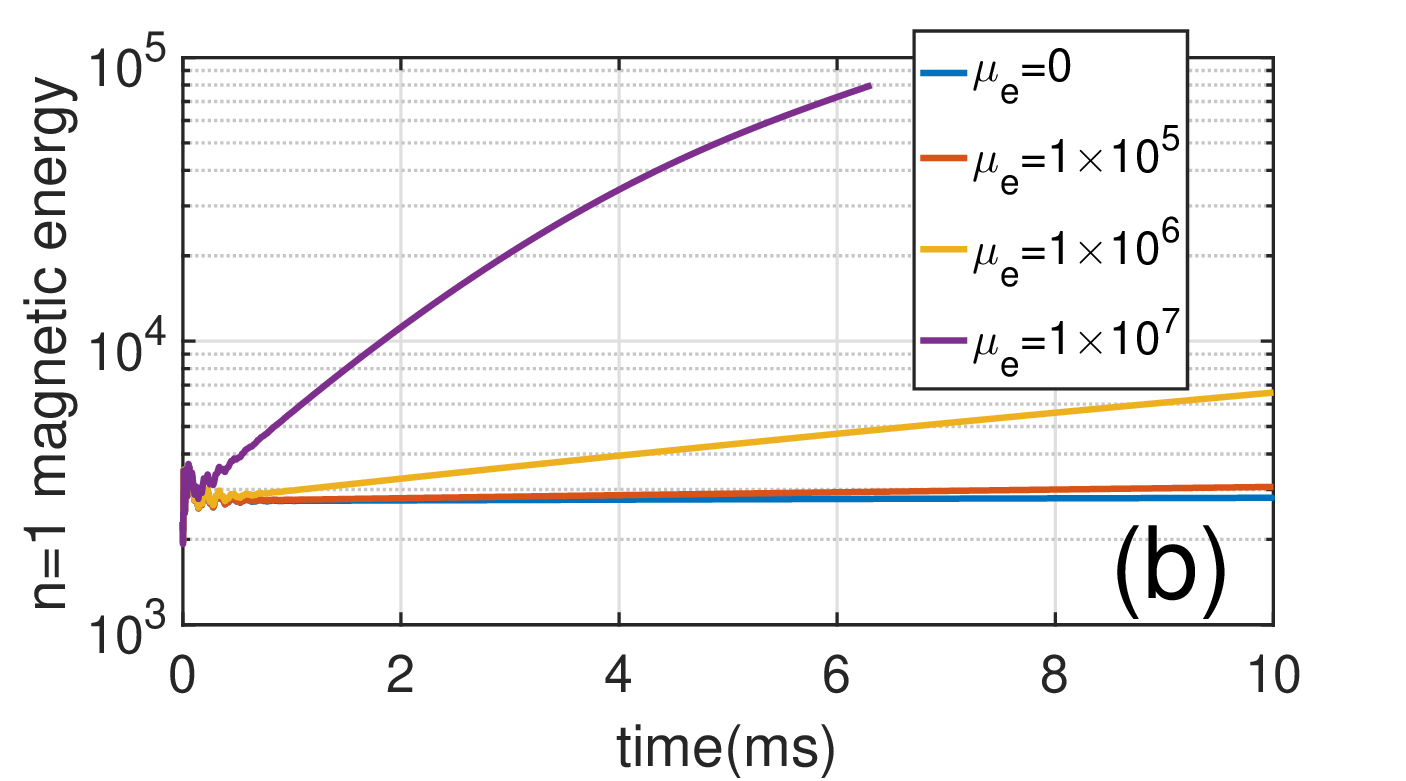}
	\end{center}
	\caption{The magnetic energy of the $n=1$ component of perturbation as a function of time with various (a) Lundquist numbers ($\mu_e=0$) and (b) neoclassical viscosity coefficient $\mu_e$ values ($S\sim10^8$) in presence of a pre-existing $2/1$ island.}
	\label{fig: stability and NTM model}
\end{figure}

Fig. \ref{fig: stability and NTM model}(a) shows that the equilibrium is unstable to the $n=1$ mode in the low Lundquist number regime ($S\sim10^4$), and can be regarded to be marginal stable in the high Lundquist number regime ($S>10^6$), which results from the largely reduced mode growth rate in the high Lundquist number regime and the stabilization effect of the toroidal curvature. Therefore, the simulations in the following sections are mainly implemented with the large Lundquist number ($S\sim10^7$ and $S\sim10^9$) unless noted otherwise.

The simulation results presented in Fig. \ref{fig: stability and NTM model}(b) demonstrate the NTM onset and growth due to the inclusion of the neoclassical electron viscous stress tensor. A pre-existing $m=2/n=1$ island is added to the equilibrium at the beginning of simulations (where $m$ is the poloidal mode number), and the mode is marginally stable before the neoclassical closure is introduced (i.e. when $\mu_e=0$). The $2/1$ island becomes visibly unstable when $\mu_e$ is sufficiently large and the island grows more rapidly with the enhanced coefficient $\mu_e$, which is due to the increase of the perturbed bootstrap current brought by the neoclassical electron viscous stress in the extended Ohm's law (\ref{eq:Ohm law}).

\section{Impurity radiation induced seed island}
\label{Sec:impurity radiation induced seed}
\begin{figure}[ht]
	\begin{center}
		\includegraphics[width=0.45\linewidth]{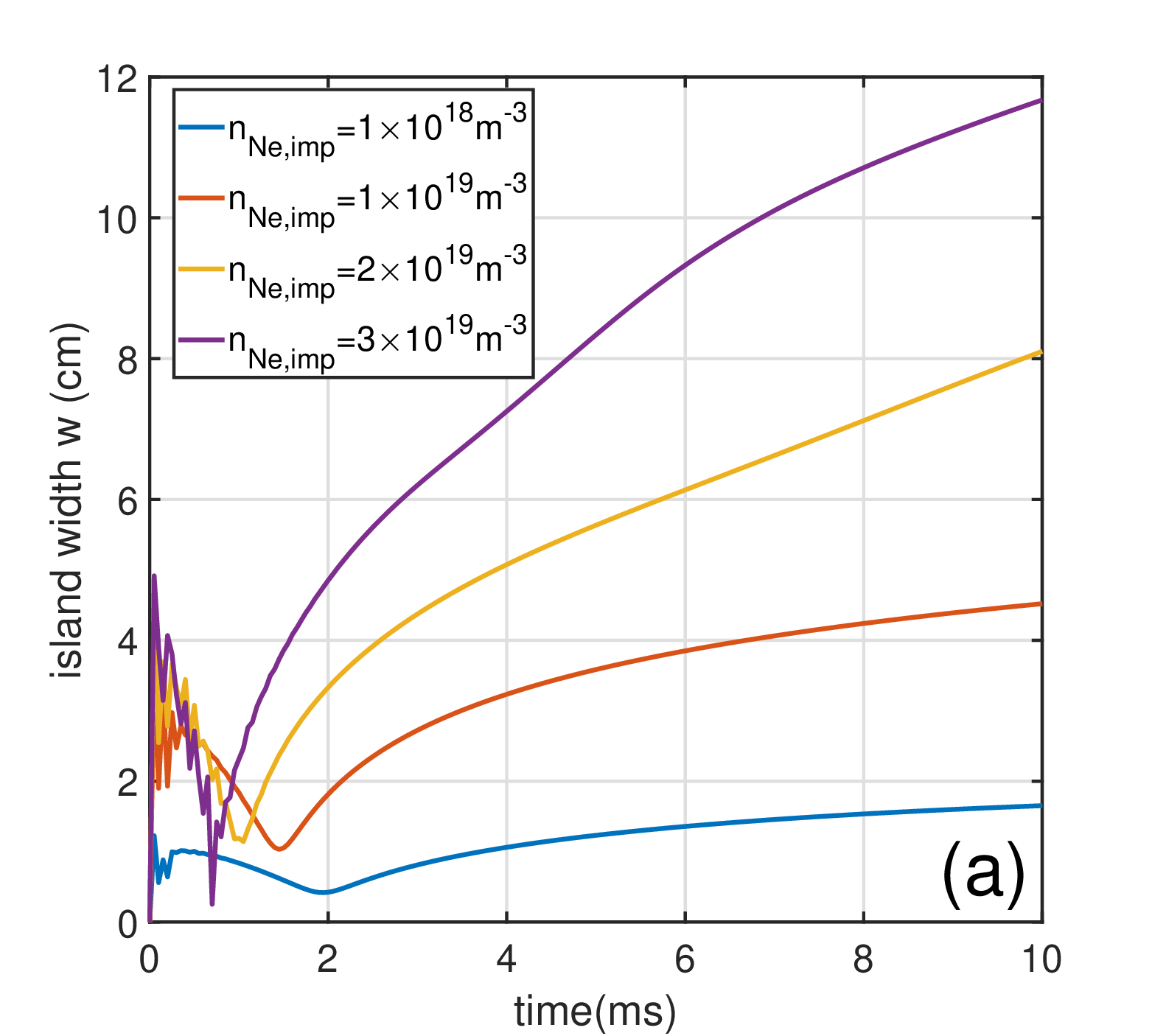}
		\includegraphics[width=0.45\linewidth]{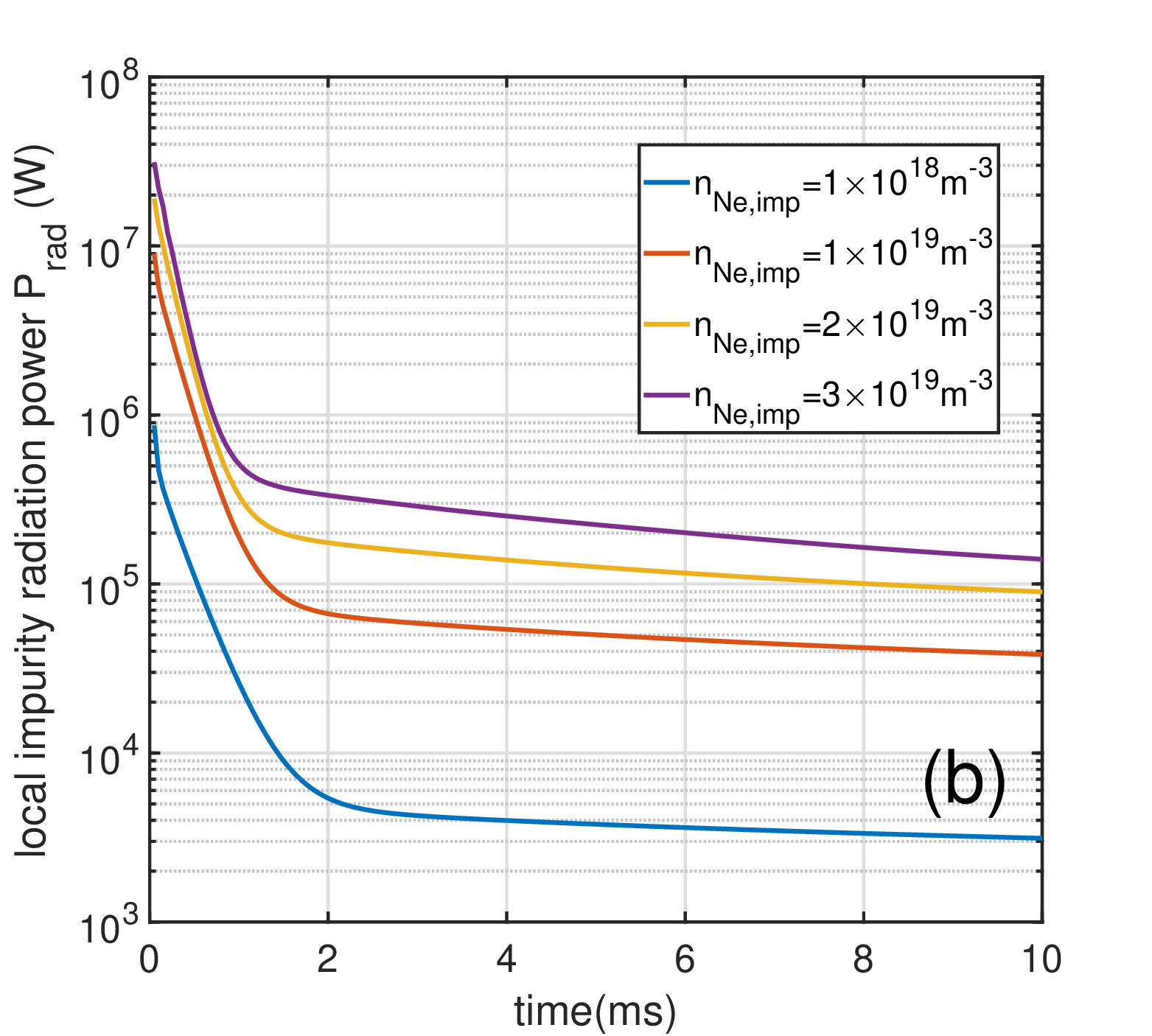}
	\end{center}
	\caption{(a) The island width of the $2/1$ mode and (b) the local impurity radiation power on the $q=2$ surface as functions of time with various impurity density levels $n_{Ne,imp}$, where the Lundquist number $S\sim10^7$.}
	\label{fig: TM island width and prad}
\end{figure}

First, we focus on the simulations with different local impurity density levels $n_{Ne,imp}$ in absence of the neoclassical closure (Figs. \ref{fig: TM island width and prad}). The equilibrium is marginally stable in the large Lundquist number regime ($S\sim10^7$) as mentioned above. The impurity radiation induced resistive magnetic island width and growth rate are proportional to the impurity density level (Fig. \ref{fig: TM island width and prad}a), as well as the correspondingly local impurity radiation power (Fig. \ref{fig: TM island width and prad}b), the island width is calculated using $w=4\sqrt{rqB_r^{2/1}/mq'B_{\theta0}}$ with all values determined at the local rational surface, where $r$ is the radius of the rational surface, $q$ is the safety factor, $B_r^{2/1}$ is the $m=2/n=1$ radial component of the perturbed magnetic field and $B_{\theta0}$ is the equilibrium poloidal magnetic field. Specifically, the island almost saturates at a very small size with low impurity density level.

\begin{figure}[ht]
	\begin{center}
		\includegraphics[width=0.45\linewidth]{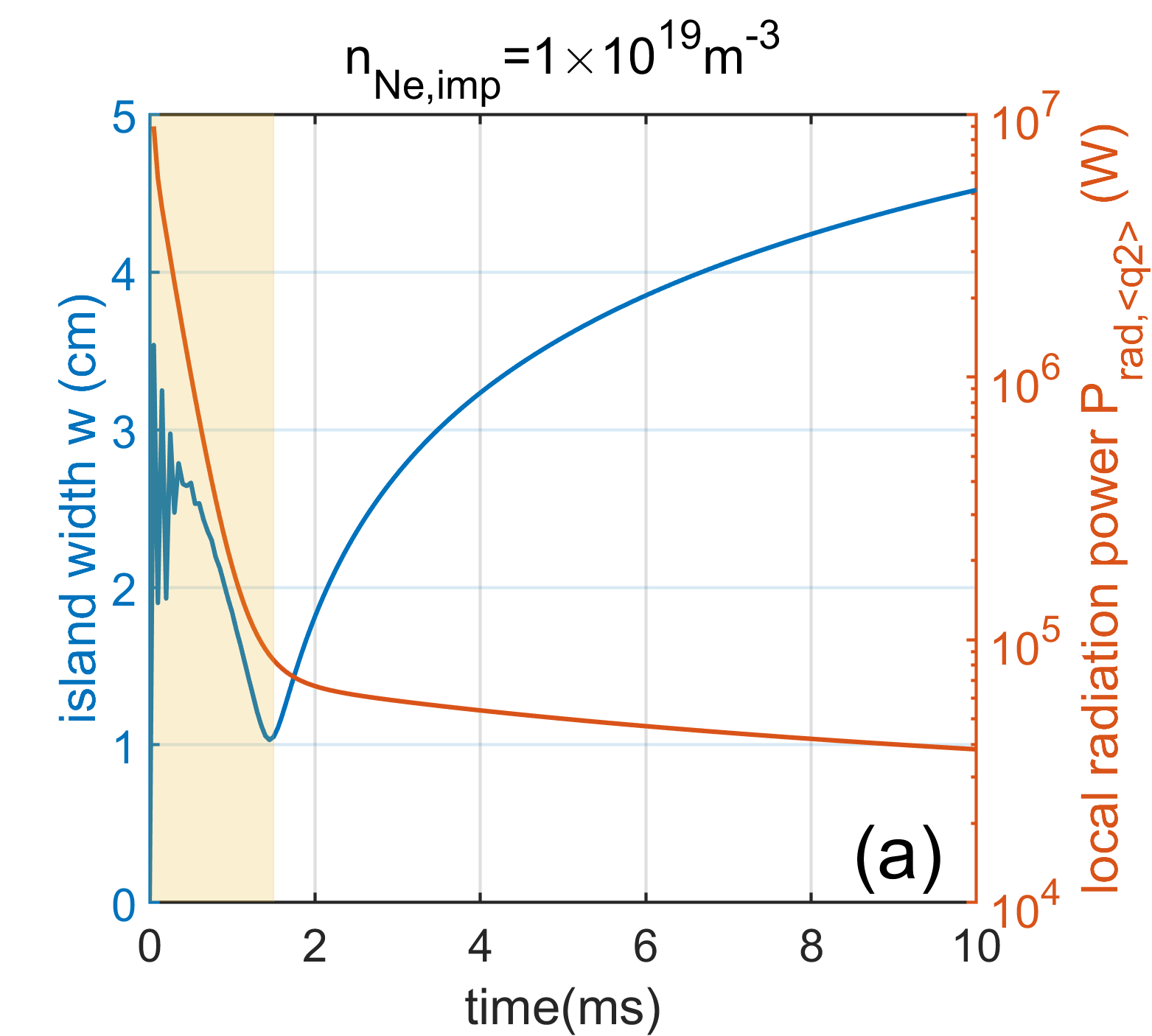}
		\includegraphics[width=0.45\linewidth]{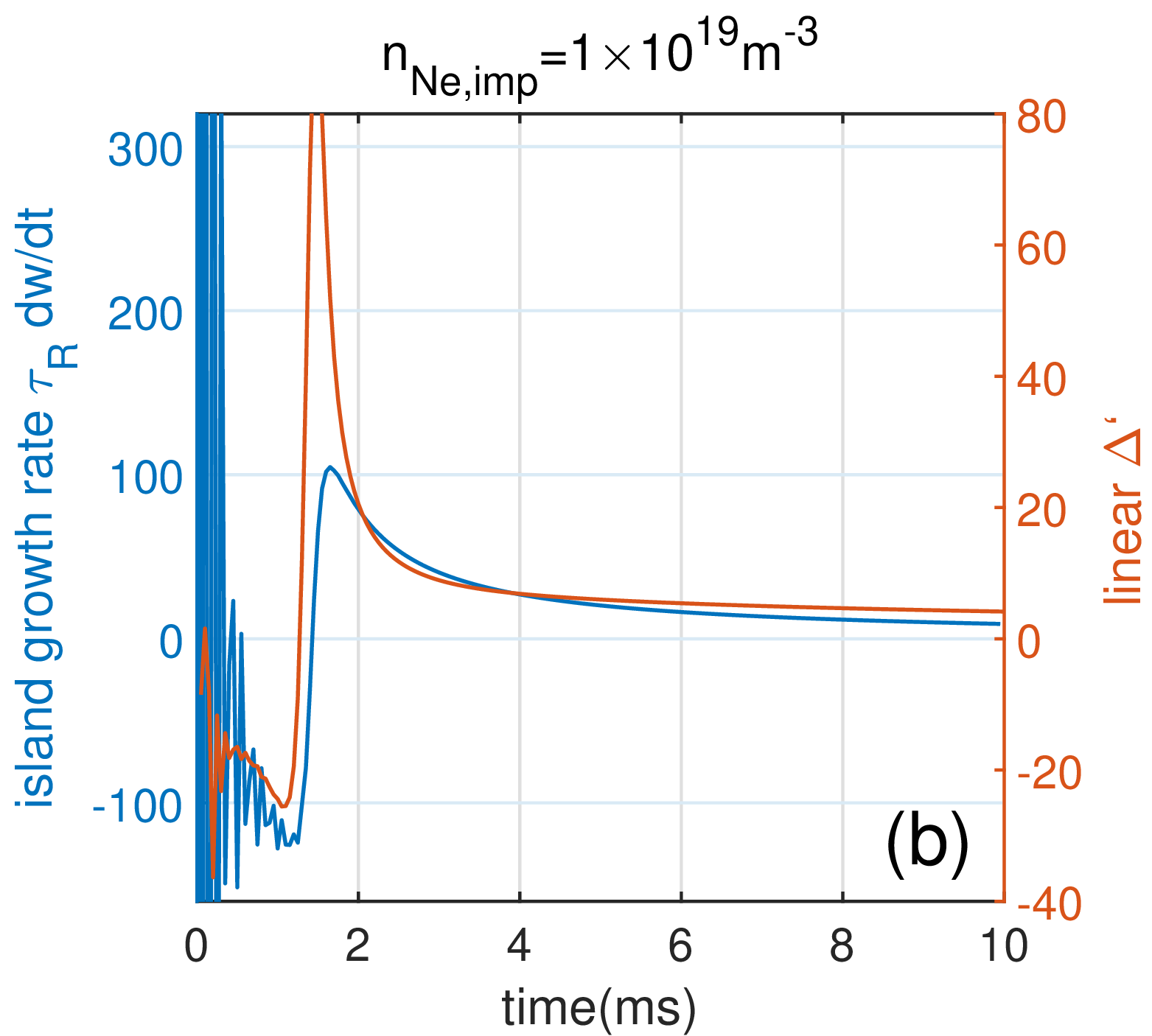}
	\end{center}
	\caption{(a) The island width of the $2/1$ mode (blue) and the local impurity radiation power on the $q=2$ surface (orange), and (b) the island growth rate of the $2/1$ mode (blue) and the linear tearing stability parameter $\Delta{'}$ as functions of time, where the impurity level $n_{Ne,imp}=1\times10^{19}m^{-3}$ and the Lundquist number $S\sim10^7$.}
	\label{fig: TM ne1e19 island-dwdt-prad}
\end{figure}

In the following analysis, the small island growth case, i.e. $n_{Ne,imp}=1\times10^{19}$m$^{-3}$, is studied in detail to elucidate the effects of local impurity radiation cooling on the seed island growth. The large impurity radiation power peak at the beginning of the simulation is dominated by the line radiation from the neutral source, which then decays quickly with the increase of impurity ionization, as shown in the yellow shadow region of Fig. \ref{fig: TM ne1e19 island-dwdt-prad}(a). The radiation peak induces drastic initial perturbations to the plasma equilibrium on the thermal transport time scale $\tau_{th}=(2\pi R_0)^2/\kappa_{\parallel}\sim 10^{-7}$s, as indicated by the oscillation in the island width ($t=0-0.5$ms), whose period is much shorter than the resistive diffusion time scale $\tau_R=\mu_0a^2/\eta\sim10^2$s. The initial oscillation in island width soon turns into a decline as the radiation power quickly decays. The tearing mode becomes the dominant instability after $t>1.5$ms following the end of the initial decay, and the magnetic island growth rate $dw/dt$ increases rapidly to a maximum before decreasing gradually towards a steady value (Fig. \ref{fig: TM ne1e19 island-dwdt-prad}b). Meanwhile, the tearing stability parameter $\Delta^{'}$ at the resonant surface evolves in a similar manner, which surges towards to positive maximum from negative sharply and drops to a slow decay afterwards. Thus the impurity radiation induced seed island onset and growth is consistent with the Rutherford model $\tau_R {\rm d}w/{\rm d}t \sim \Delta^{'}$ most of time, except around the peaking moment when other non-inductive contributions to the helical current layer at resonant surface also become substantial, such that $\Delta^{'}$ should be replaced with $\Delta^{'}-\Delta^{'}_c$, where $\Delta^{'}_c$ is the Glasser resistive interchange correction due to toroidal curvature effects \cite{Glasser1975,Glasser_1976}.
The surge of $\Delta^{'}$ term to extremely large positive value at the beginning of nonlinear mode growth is very similar to the spontaneous NTM onset proposed by Brennan et al \cite{Brennan2002,Brennan2003}. However, the dynamics of the $\Delta^{'}$ term in our simulations arises from the local helical current perturbation induced by the local impurity radiation cooling, and the value of $\beta_N$ decreases in the presence of impurity radiation rather than approaching to the ideal stability limit. In addition, the local enhanced resistivity due to the radiative cooling reinforces the tearing mode growth as well.

\begin{figure}[ht]
	\begin{center}
		\includegraphics[width=0.55\linewidth]{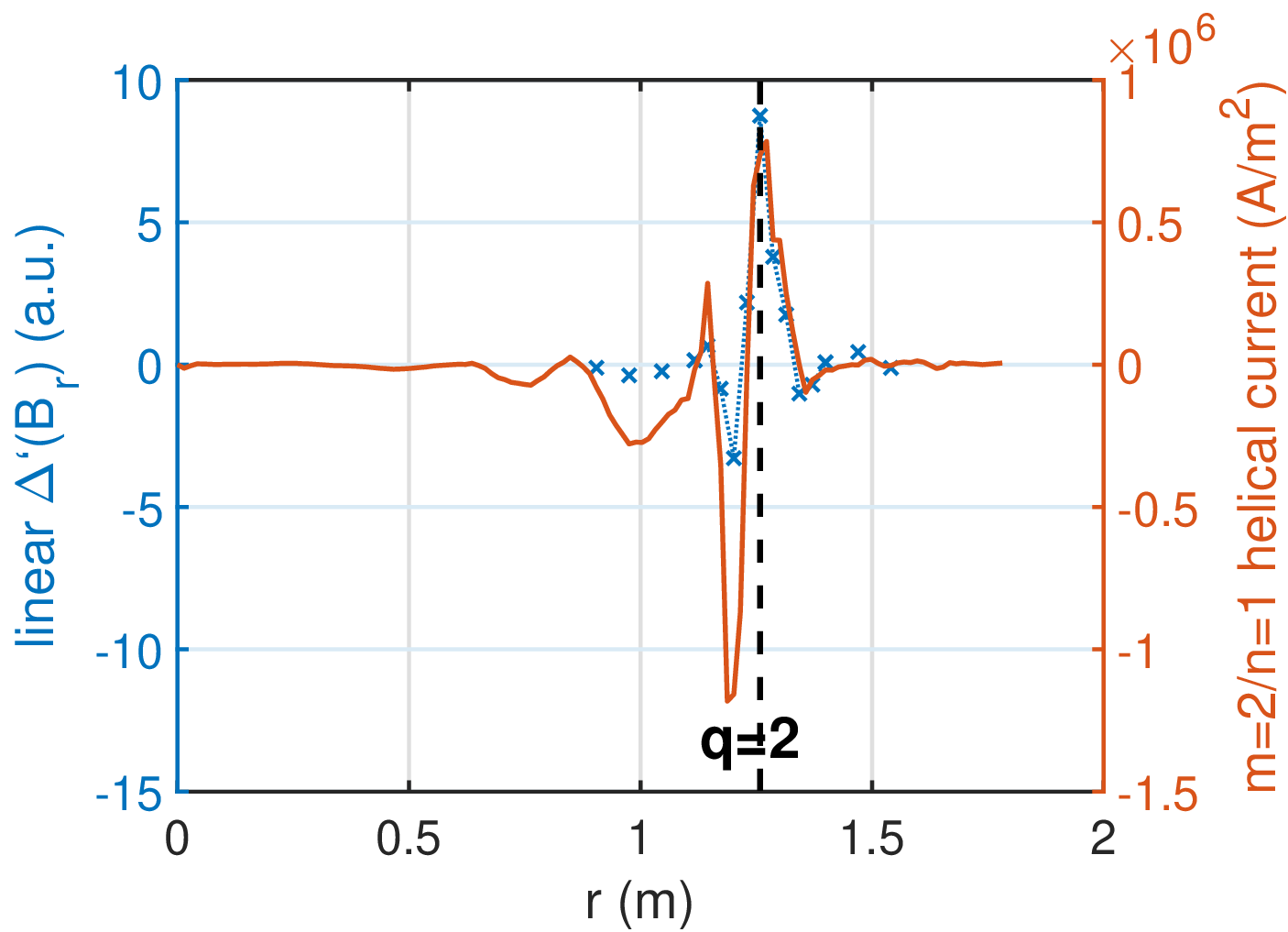}
	\end{center}
	\caption{The radial profile of the $2/1$ component of helical current density parallel to the equilibrium magnetic field (orange) and the discrete values of the linear $\Delta^{'}$ (blue) around the rational surface, the blue dotted line shows the variation of the $\Delta^{'}$	and the $q=2$ surface is indicated by the vertical black dashed line, where the impurity level $n_{Ne,imp}=1\times10^{19}$m$^{-3}$, the Lundquist number $S\sim10^7$ and $t=2.5$ms.}
	\label{fig: TM ne1e19 delta-Jh}
\end{figure}

\begin{figure}[ht]
	\begin{center}
		\includegraphics[width=0.45\linewidth]{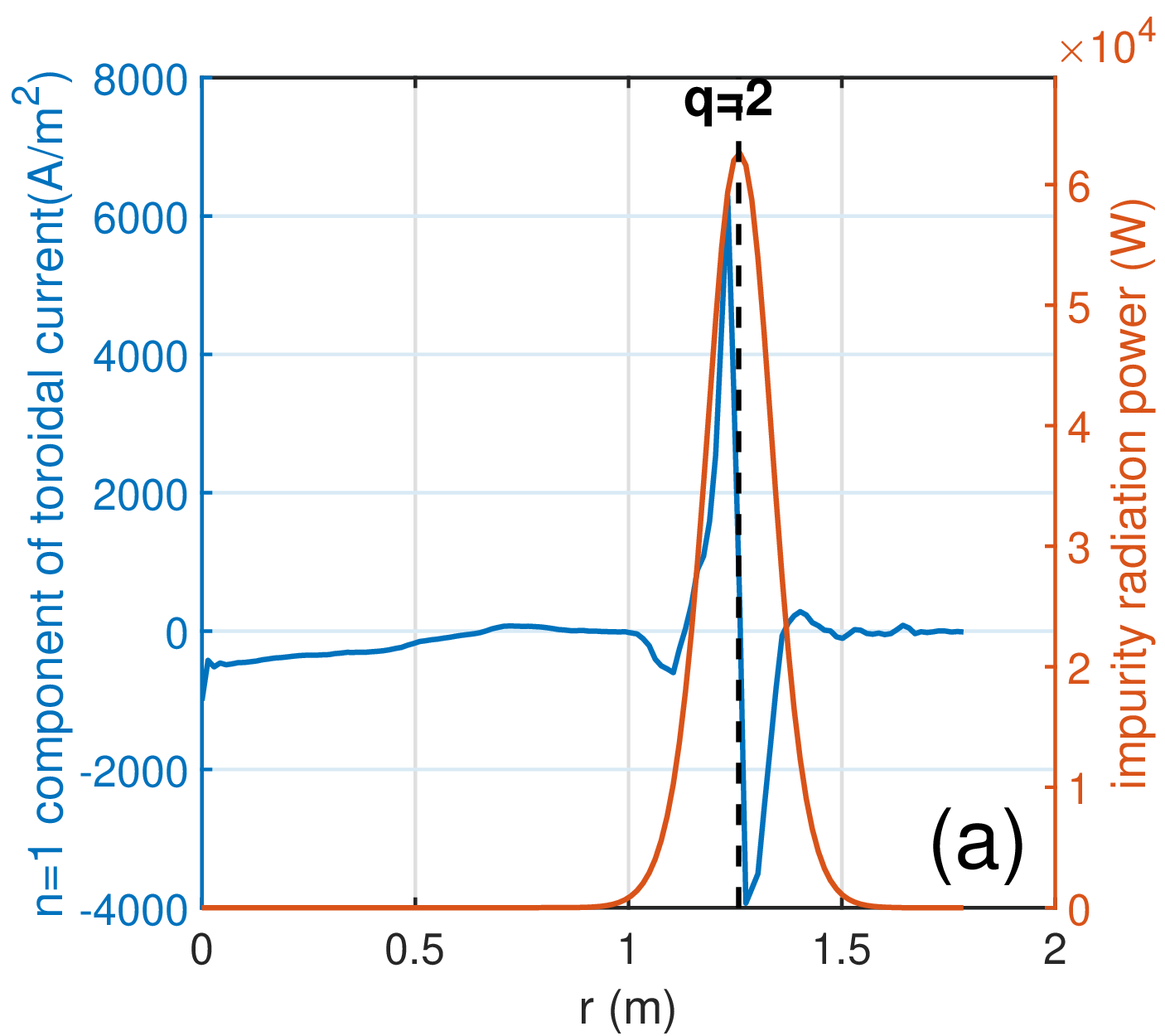}
		\includegraphics[width=0.45\linewidth]{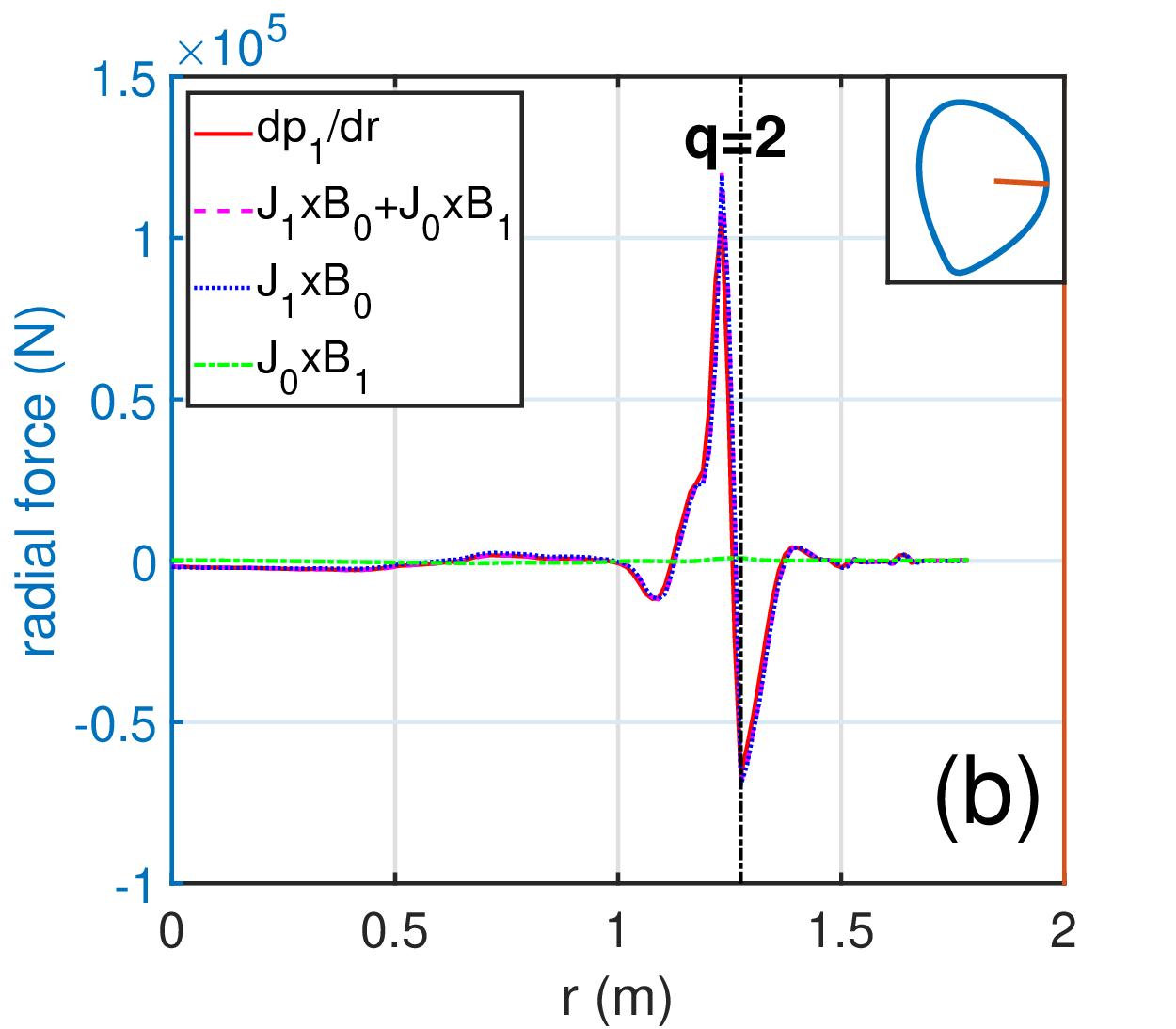}
	\end{center}
	\caption{The radial profiles of (a) the $n=1$ component of toroidal current density (blue) and the impurity radiation power (orange), and (b) the radial gradient of the $n=1$ component of pressure ${\rm d}p_1/{\rm d}r$ (red solid), the $n=1$ radial components of Lorentz force $\vec{J}_1\times \vec{B}_0+\vec{J}_0\times \vec{B}_1$ (magenta dashed), $\vec{J}_1\times \vec{B}_0$ (blue dotted), and $\vec{J}_0\times \vec{B}_1$ (green dash-dot), where $\vec{J}_0$ ($\vec{B}_0$) and $\vec{J}_1$ ($\vec{B}_1$) refer to the $n=0$ and $n=1$ component of the current density (magnetic field) respectively, where the impurity level $n_{Ne,imp}=1\times10^{19}m^{-3}$ and the Lundquist number $S\sim10^7$.}
	\label{fig: TM ne1e19 n1 component force balance}
\end{figure}

In particular, the linear tearing stability parameter is measured on the rational surface as $\Delta^{'}=[{\rm d}B_r^{2/1}/{\rm d}r(r_s^+)-{\rm d}B_r^{2/1}/{\rm d}r(r_s^-)]/B_r^{2/1}(r_s)$, where $r_s$, $r_s^+$, $r_s^-$ refers to the radial location of the rational surface, the right hand side and the left hand side of $r_s$, respectively, and the $\Delta^{'}$ term is measured at every time step based on the dynamic plasma profiles in the simulations.
The location of the rational surface is determined from the safety factor profile of the initial equilibrium. In order to avoid difficulties associated with identifying the layer width, we compute the jump in $B_r^{'}$ at the finite element node locations on either side of the rational surface. We find that this approximation of $\Delta^{'}$ works well in the small island limit, but we expect it to break down in the large island limit where the nonlinear form $\Delta^{'}(w)$ applies. Essentially, the tearing stability parameter is a simplified representation of the Ampere's law, i.e. $\Delta^{'}\psi\simeq\int\nabla^2\psi {\rm d}r \simeq \int_{r_s^-}^{r_s^+}\delta J {\rm d}r$, where $\psi$ is the perturbed magnetic flux, and $\delta J$ refers to the helical current density perturbation parallel to the equilibrium magnetic field within the tearing layer. As shown in Fig. \ref{fig: TM ne1e19 delta-Jh}, after dividing the region around the rational surface into several layers with each enclosed by adjacent grid points, the corresponding discrete values of $\Delta^{'}$ are able to represent the helical current density perturbations within each of those layers around the rational surface, except for the initial $0-1.5$ms when the drastic impurity radiation induced perturbations dominate.
The measured $\Delta^{'}$ term includes contributions from
the current perturbations induced by the local impurity radiation cooling (Fig. \ref{fig: TM ne1e19 n1 component force balance}a), primarily through the modification of the local pressure profile around the rational surface (Fig. \ref{fig: TM ne1e19 n1 component force balance}b). The perturbed pressure gradient peaks at both sides of the rational surface, and gives rise to the modifications of the $\Delta^{'}$ from the plasma profiles in the ideal region outside the resistive layer. Besides, the positive pressure gradient perturbation $dp_1/dr$ inward of the rational surface, i.e. $r<r_s$, and on the contrary a negative $dp_1/dr$ outward, i.e. $r>r_s$, leads to a flat or even hollow pressure profile across the rational surface. Based on the modified local static force balance ${\rm d}p_1/{\rm d}r = J_1 \times B_0 + J_0 \times B_1$, where the second part on the right hand side can be neglected (Fig. \ref{fig: TM ne1e19 n1 component force balance}b), the $n=1$ component of the perpendicular current perturbation mainly consists of the diamagnetic current in the outer ideal radiative cooling region and can be approximated with $J_{1\perp} \simeq \frac{1}{B_0} {\rm d}p_1/{\rm d}r$, which in turn contributes to the parallel current perturbation $J_{1\parallel}$ through the quasi-neutrality condition $\nabla\cdot\vec{J}_1=0$. Thus, in the limit of small island size, the seed island growth can be well approximated by the Rutherford model
\begin{equation}
	\tau_R\frac{{\rm d}w}{{\rm d}t} \simeq \Delta^{'} -\Delta^{'}_c,
	\label{eq:MRE for the seed island}
\end{equation}
where the $\Delta^{'}$ is mainly driven by the modification of the plasma profile in the outside region of the resistive layer, as a result of the helical current perturbation induced by the local impurity radiation cooling, and is matched with the response from the inductive current perturbation that gives rise to the island growth inside the resistive layer.
$\Delta^{'}_c$ represents the diamagnetic correction due to the stabilization effect of toroidal curvature inside the resistive layer at the early linear phase \cite{Glasser1975}.

\begin{figure}[ht]
	\begin{center}
		\includegraphics[width=0.45\linewidth]{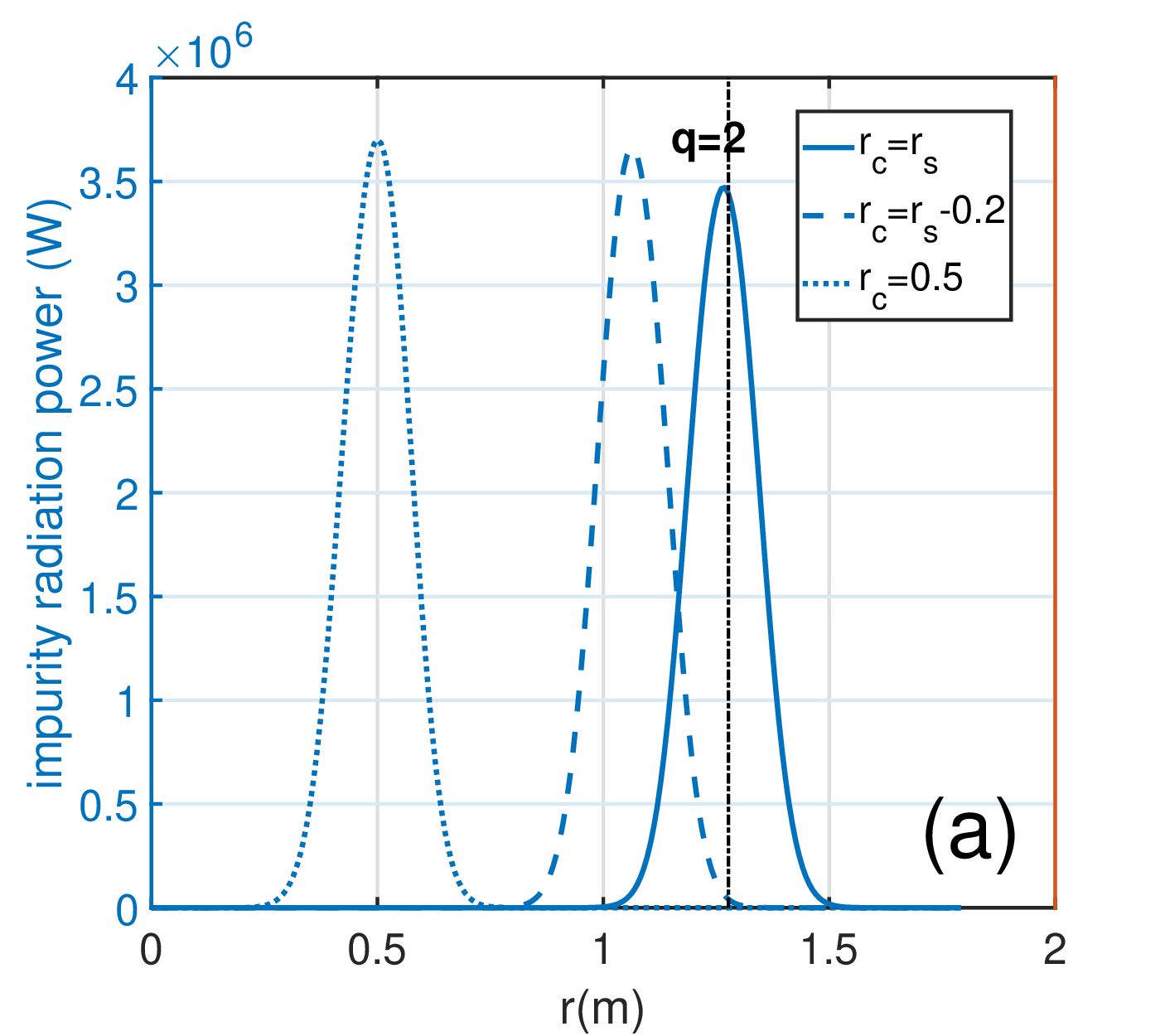}
		\includegraphics[width=0.45\linewidth]{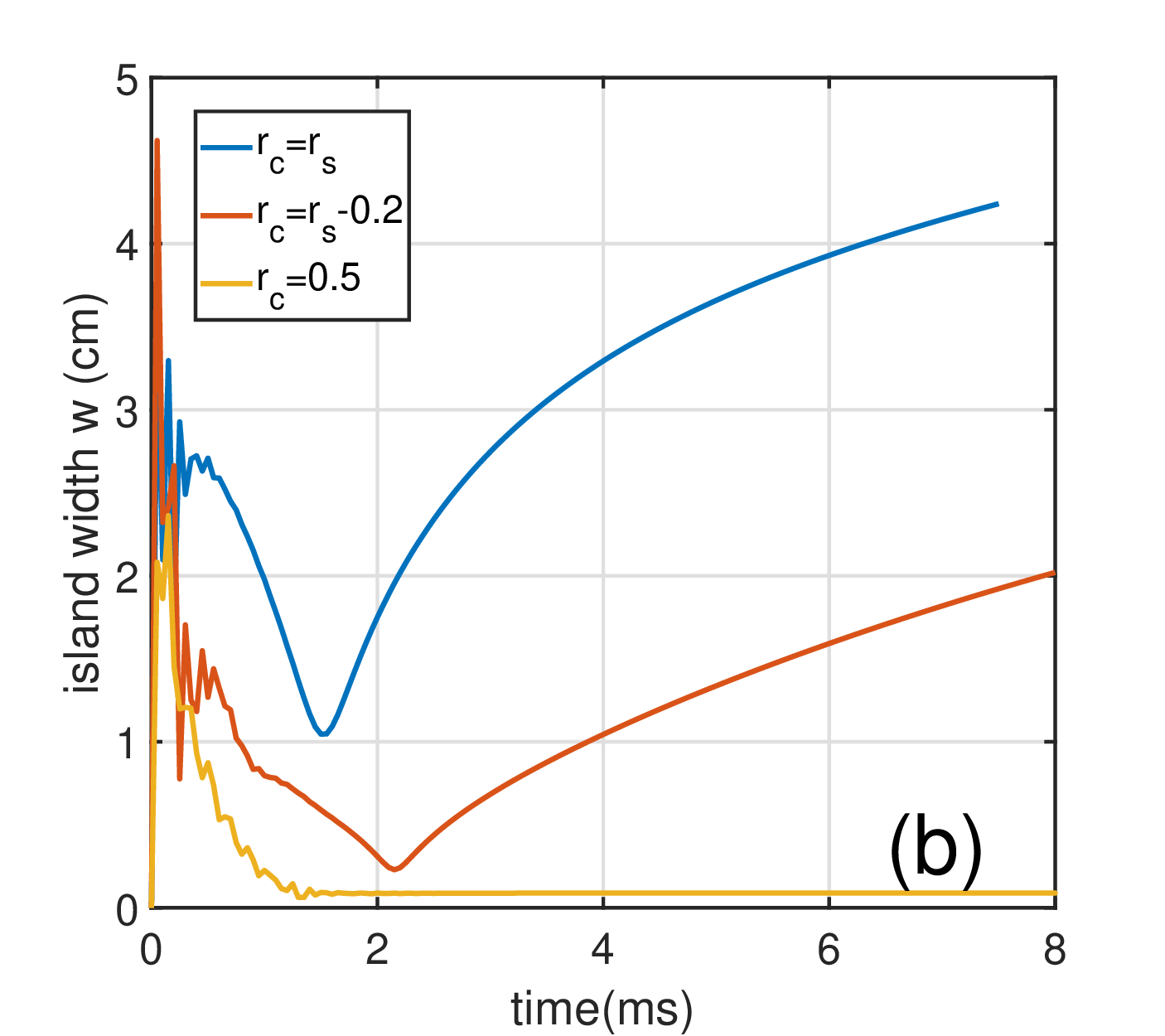}
	\end{center}
	\caption{(a) The impurity radiation power profiles at various cooling locations $r_c$, and $r_s$ refers to the location of the $q=2$ rational surface. (b) The island width of $2/1$ mode as functions of time with various $r_c$ values, where the impurity level $n_{Ne,imp}=1\times10^{19}m^{-3}$ and the Lundquist number $S\sim10^7$.}
	\label{fig: TM ne1e19 cooling location}
\end{figure}

\begin{figure}[ht]
	\begin{center}
		\includegraphics[width=0.45\linewidth]{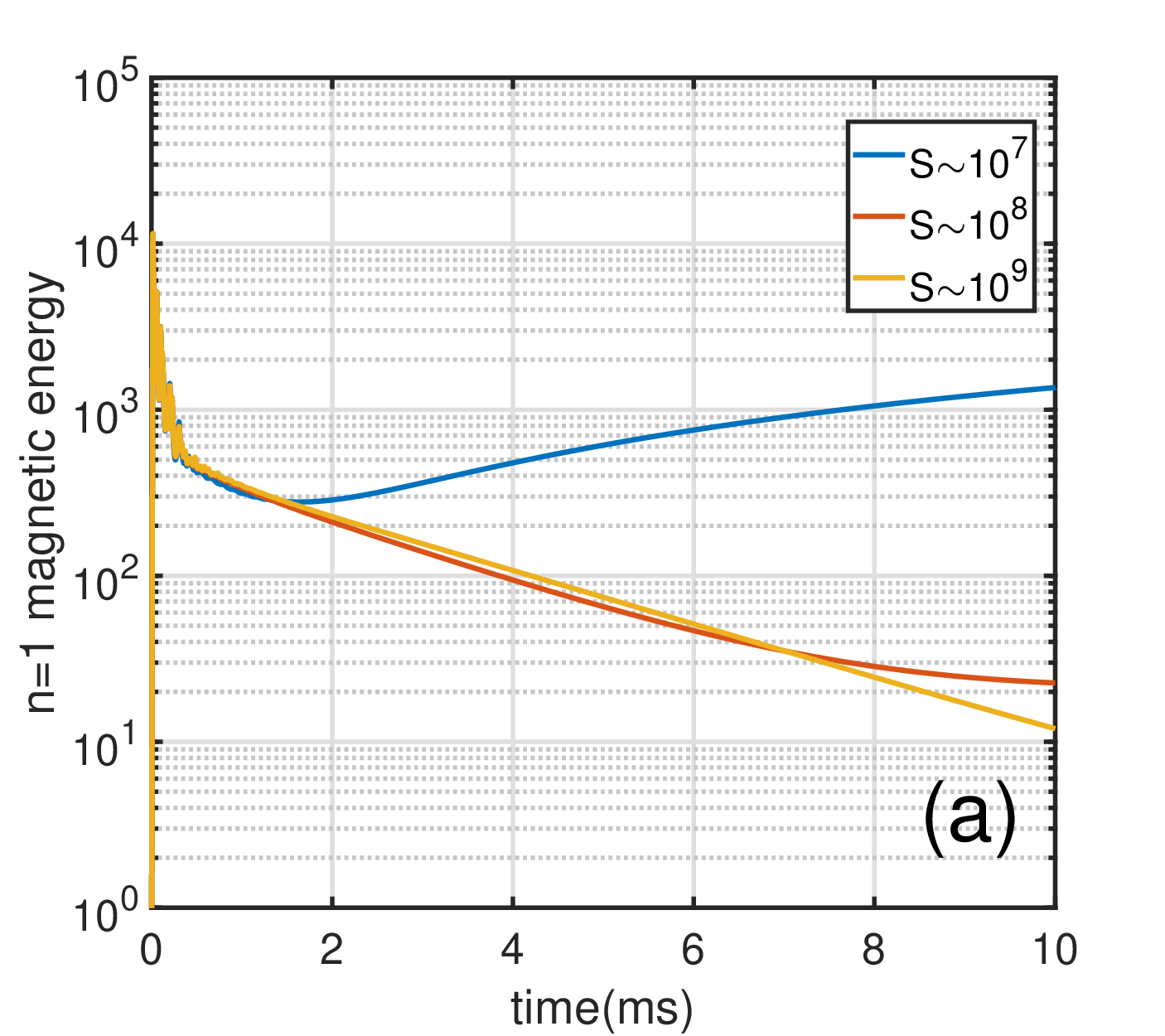}
		\includegraphics[width=0.45\linewidth]{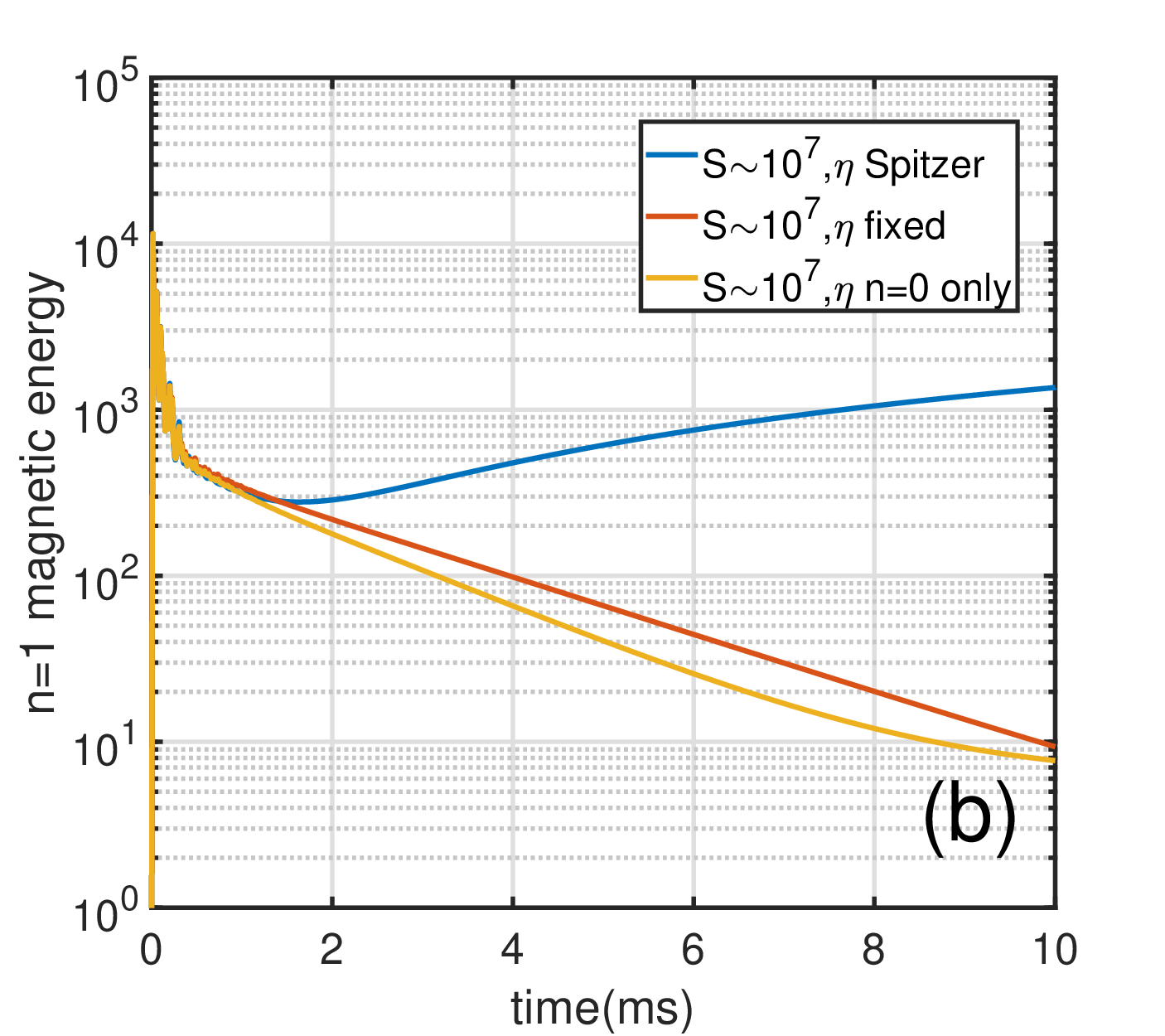}
	\end{center}
	\caption{The $n=1$ component of the magnetic energy as a function of time with various (a) Lundquist number $S$ regimes and (b) plasma resistivity $\eta$ models: the full Spitzer model (blue), the uniform resistivity model (red), and the Spitzer model with symmetric component only (yellow), where the impurity level $n_{Ne,imp}=1\times10^{19}m^{-3}$ and the Lundquist number $S\sim10^7$.}
	\label{fig: TM ne1e19 plasma resistivity}
\end{figure}

This model can be confirmed by the simulation results with different locations of impurity radiation cooling (Figs. \ref{fig: TM ne1e19 cooling location}). As the radiative cooling source is located far away from the rational surface (the $q=2$ surface), the island width and growth rate are both reduced and even decay away.
Meanwhile, the onset and growth rates of the seed island is proportional to the plasma resistivity, which is characteristic of the resistive tearing mode, as shown in Fig. \ref{fig: TM ne1e19 plasma resistivity}(a), and the initial perturbations decay much longer over time before the onset of island growth in the large Lundquist number regime.
In particular, the helical component of the plasma resistivity is found crucial to the seed island growth (Fig. \ref{fig: TM ne1e19 plasma resistivity}b), the rapid mode growth is only found with the full Spitzer resistivity model but not with either the constant resistivity or the Spitzer model including the axisymmetric part only, which presumably is due to the large time scale of resistive diffusion and the lack of corporation between the resonant components of the plasma resistivity and the helical current perturbation, respectively.
Our simulation results agree with the report in Ref. \cite{Muraglia_2021}, which demonstrates the numerical island dynamic matches the theory model well only in the small island size region, and more sophisticated physics are involved with the consideration of finite island width.

\begin{figure}[ht]
	\begin{center}
		\includegraphics[width=0.65\linewidth]{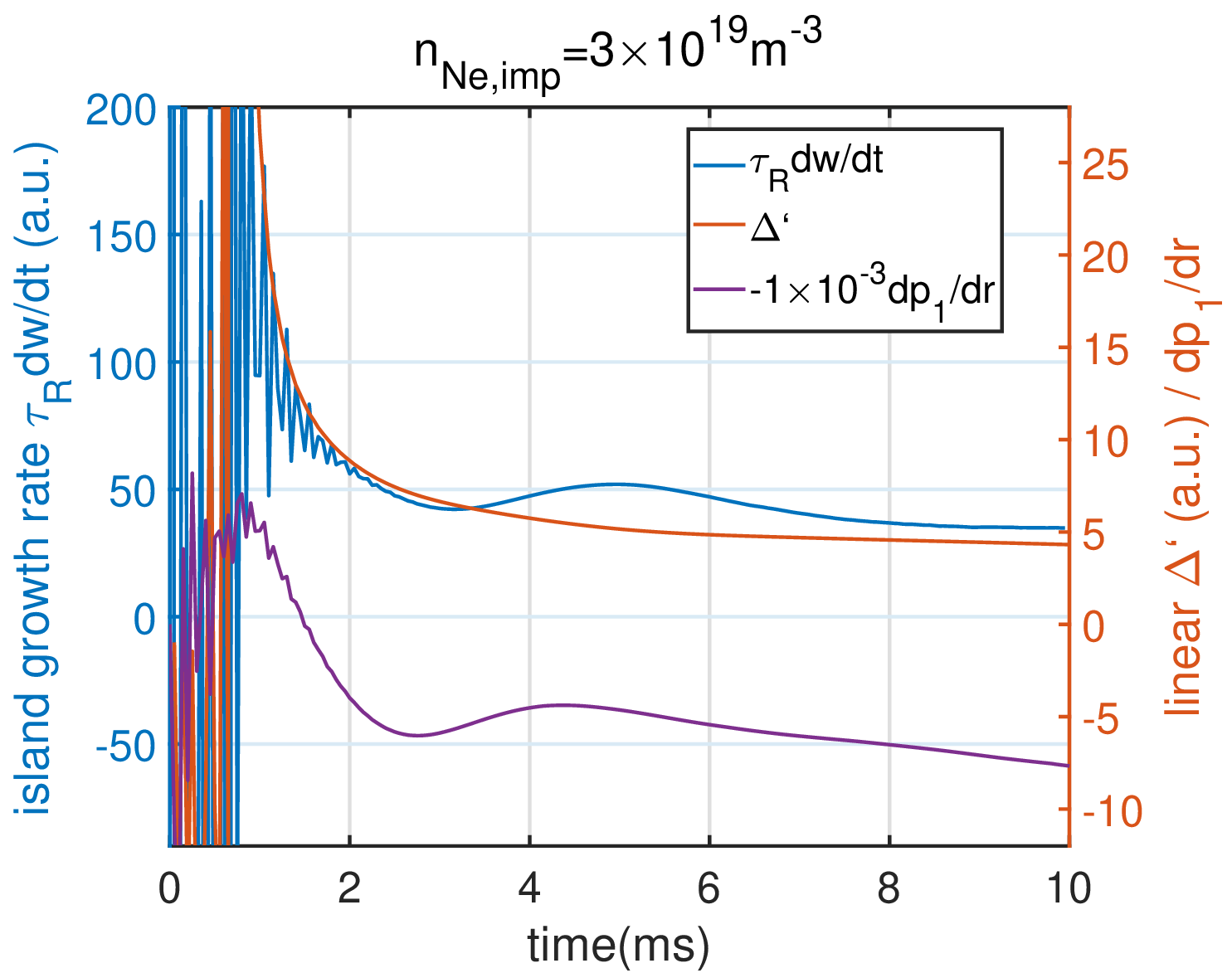}
	\end{center}
	\caption{The island growth rate of the $2/1$ mode (blue), the linear tearing stability parameter $\Delta^{'}$ (orange), and the $n=1$ component of pressure gradient (purple) as functions of time, where the impurity level $n_{Ne,imp}=3\times10^{19}m^{-3}$ and the Lundquist number $S\sim10^7$.}
	\label{fig: TM ne3e19 island growth}
\end{figure}

For the nonlinear mode growth in the later stage of large island growth and radiation power, a hump appears in the mode growth rate as shown in Fig. \ref{fig: TM ne3e19 island growth}. The nonlinear tearing stability parameter $\Delta^{'}(w)$ term with the correction of finite island width should be adopted, however, the exact form of $\Delta^{'}(w)$ could be complicated due to the asymmetry of magnetic island structure in toroidal geometry. In addition, the non-inductive current perturbation from the pressure gradient perturbation due to the radiative cooling inside the resistive layer when the island size is larger continues to contribute to the diamagnetic corrections to the nonlinear island growth (Fig. \ref{fig: TM ne3e19 island growth}).
At this stage, extended Rutherford equation, such as those in Ref. \cite{Hegna_1999} should apply.

\section{Impurity radiation seeded nonlinear tearing mode growth}
\label{Sec:impurity radiation seeded NTM}
\begin{figure}[ht]
	\begin{center}
		\includegraphics[width=0.45\linewidth]{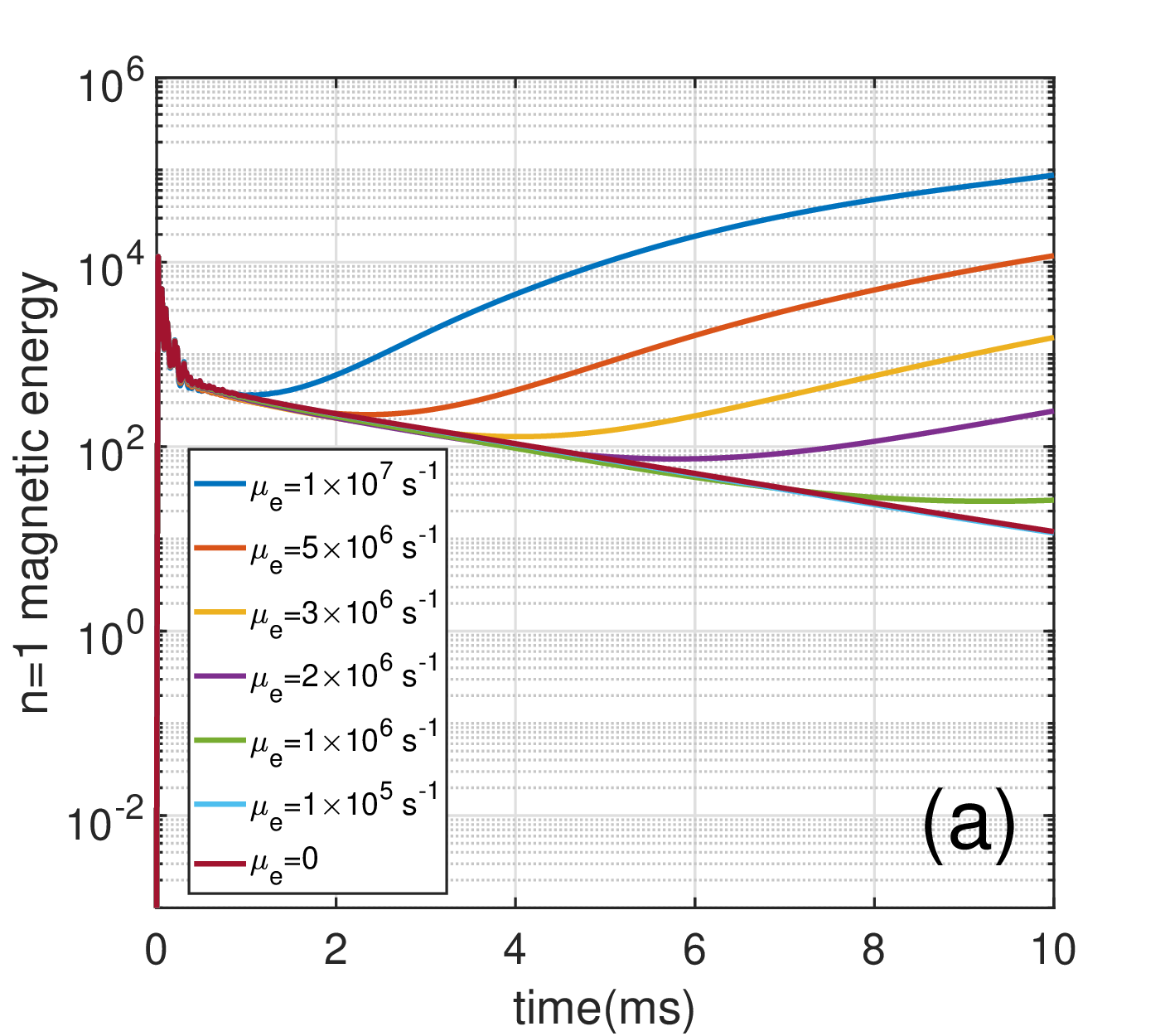}
		\includegraphics[width=0.45\linewidth]{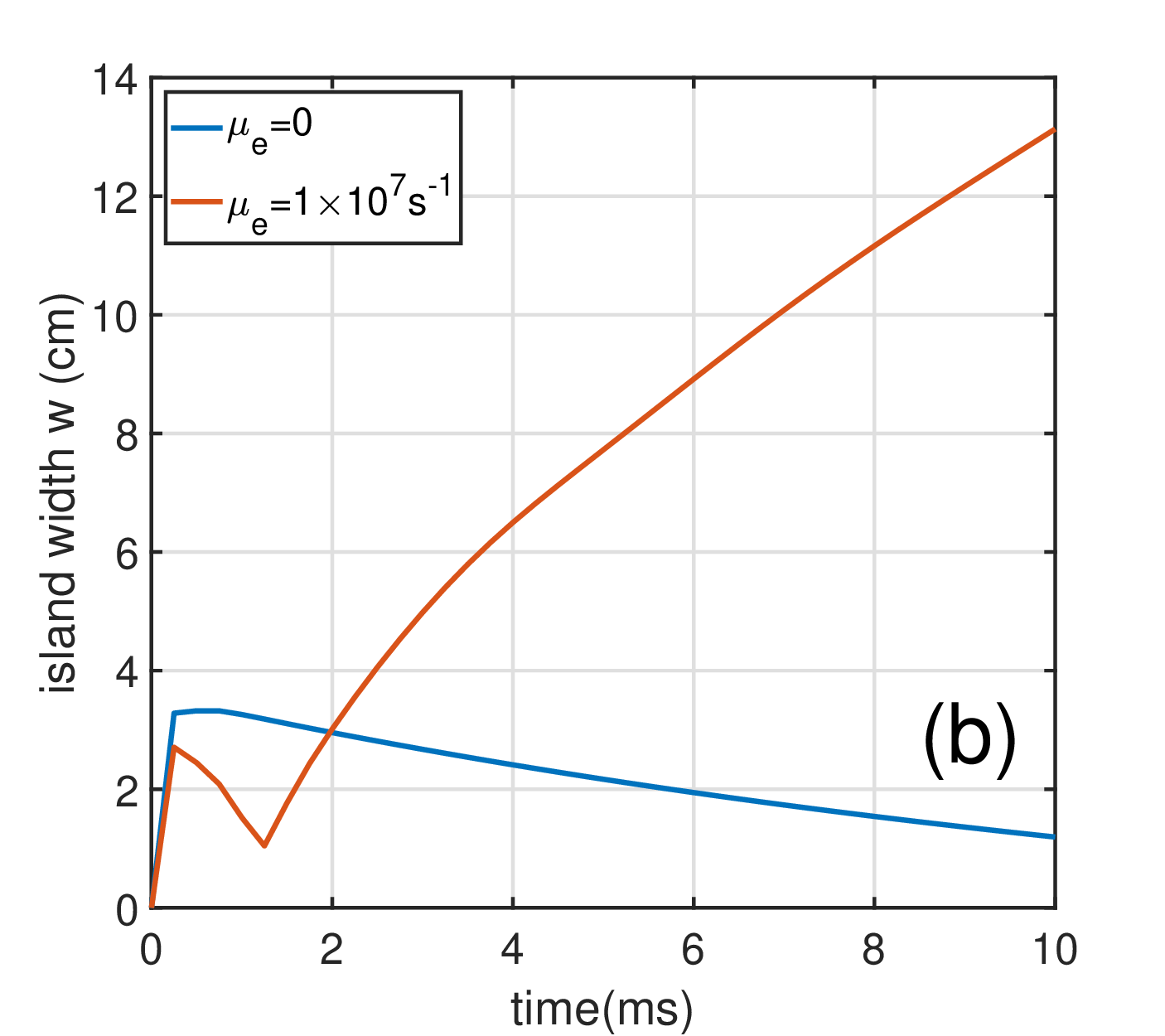}
	\end{center}
	\caption{(a) The magnetic energy of the $n=1$ component of perturbation and (b) the island width of the $2/1$ mode with various values of the coefficient $\mu_e$ as functions of time, where the impurity level $n_{Ne,imp}=1\times10^{19}m^{-3}$ and the Lundquist number $S\sim10^9$.}
	\label{fig: NTM mu-e scaling}
\end{figure}

In presence of the heuristic neoclassical closure, the growth of the impurity radiation seeded island is further driven by the perturbed bootstrap current, which originates from the neoclassical electron stress in the extended Ohm's law. As indicated by Fig. \ref{fig: NTM mu-e scaling}(a), a threshold in the coefficient $\mu_e=1\times10^6$s$^{-1}$ is identified, above which the seed island would start to grow instead of continue to decay and higher values of the coefficient $\mu_e$ lead to larger growth rates. Apparently, the inclusion of the heuristic neoclassical closure is directly responsible for such a nonlinear island growth (Fig. \ref{fig: NTM mu-e scaling}b).

\begin{figure}[ht]
	\begin{center}
		\includegraphics[width=0.45\linewidth]{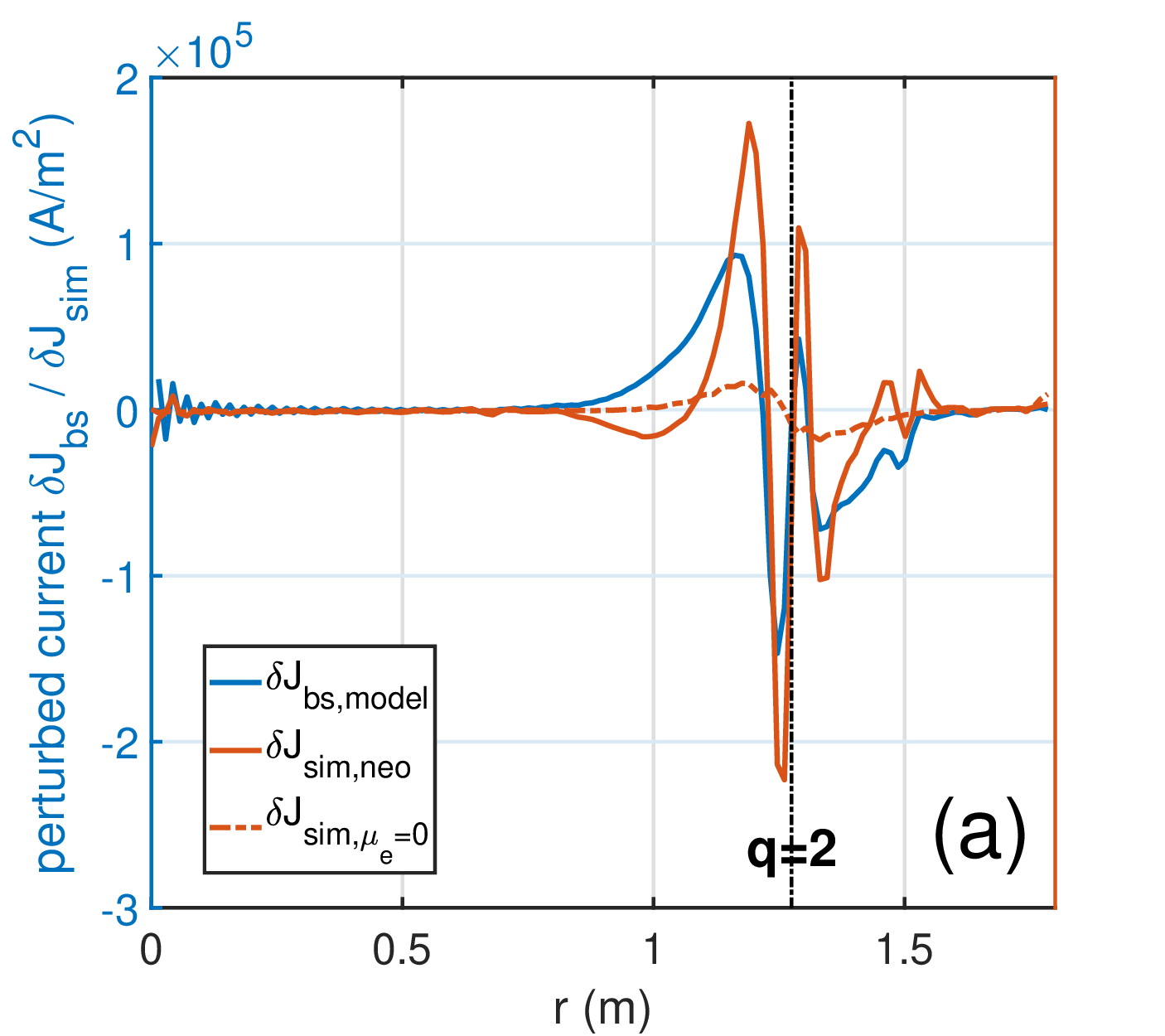}
		\includegraphics[width=0.45\linewidth]{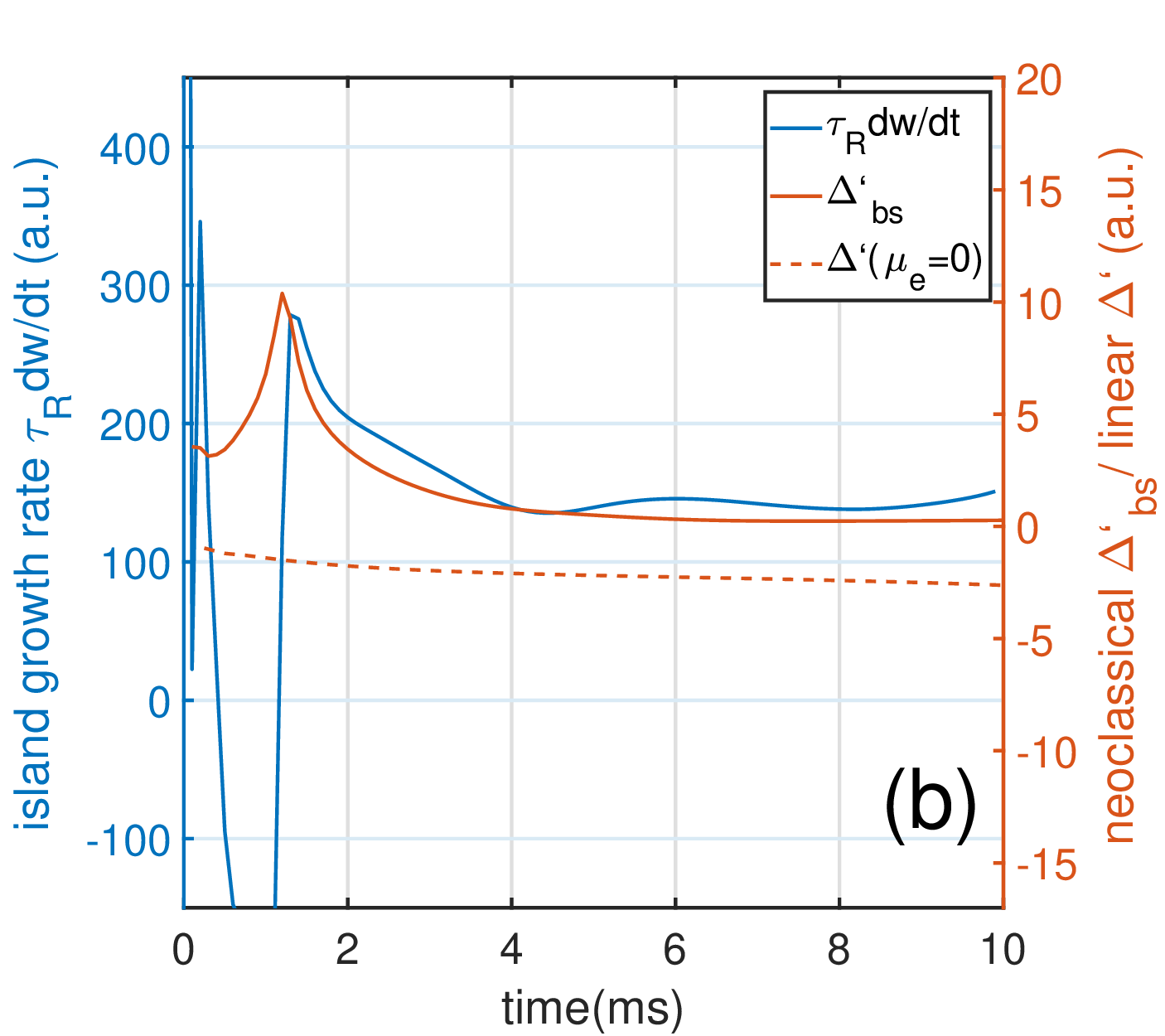}
	\end{center}
	\caption{(a) The radial profiles of the perturbed bootstrap current model $\delta J_{bs} = \epsilon^{1/2}/B_{\theta}({\rm d}\delta p/{\rm d}r)$ (blue solid line), the perturbed helical current density from the simulation results with ($\delta J_{sim,neo}$, orange solid line) and without ($\delta J_{sim,\mu_e=0}$, orange dashed line) the inclusion of the heuristic neoclassical closure, respectively. (b) The island growth rate of the $2/1$ mode (blue), the neoclassical driving term $\Delta^{'}_{bs}$ measured from the simulation results (orange solid line), and the tearing stability parameter $\Delta^{'}$ of the resistive tearing case ($\mu_e=0$, orange dashed line) as functions of time. Here the impurity level $n_{Ne,imp}=1\times10^{19}m^{-3}$, $\mu_e=1\times10^7$s$^{-1}$ and the Lundquist number $S\sim10^9$.}
	\label{fig: NTM bootstrap current calculation}
\end{figure}

A simplified theoretical model of the perturbed bootstrap current $\delta J_{bs} = \epsilon^{1/2}/B_{\theta}({\rm d}\delta p/{\rm d}r)$ is used and is generally consistent with the total current perturbations from the simulation in the NTM growth case as shown in Fig. \ref{fig: NTM bootstrap current calculation}(a), where the current perturbation in the resistive tearing mode case is much smaller in comparison to the NTM growth case, and the excessive current perturbation in the neoclassical case is attributed to the heuristic electron force $\nabla\cdot\bi{\Pi}_e/\left(en_e\right) $ in driving the bootstrap current perturbation in the extended Ohm's law. And the differences between the theory model and the simulation results come from that the current perturbations of the simulation include not only the dominant bootstrap current perturbation but also other contributions such as the perturbed  Pfirsch–Schl\"{u}ter current $\delta J_{ps}$, however, its contribution is small since $\delta J_{ps}/\delta J_{bs} \sim \epsilon^{1/2}$ \cite{Tokamak_wesson}. It is worth mentioning that the impurity density level is kept same for both the resistive and neoclassical tearing mode cases, which means the impurity radiation power level and the subsequent pressure gradient perturbation induced by radiative cooling, i.e. the perturbed Pfirsch–Schl\"{u}ter current, are comparable between the two cases, and the growth of NTM is dominated by the driving from the perturbed bootstrap current. The neoclassical driving term in the modified Rutherford equation often assumes the following simplified form in the limit of large aspect ratio
\begin{equation}
	\Delta^{'}_{bs} = \frac{\sqrt{\epsilon}}{w} \beta_p \frac{L_q}{L_p},
	\label{eq:Delta neoclassical}
\end{equation}
where $\epsilon$ is the inverse aspect ratio and $\beta_p$ the ratio of plasma thermal pressure to poloidal magnetic pressure, $L_q=q/q'$ and $L_p=p/p'$ are the scale length of the safety factor and the pressure profiles respectively, with all values evaluated on the local rational surface \cite{Hegna1998}. The neoclassical driving term $\Delta^{'}_{bs}$ measured from the simulation results using Eq. (\ref{eq:Delta neoclassical}) agrees well with the nonlinear island growth from the same simulation (Fig. \ref{fig: NTM bootstrap current calculation}b).
In addition, the tearing stability parameter $\Delta^{'}$ remains negative over time in the resistive tearing mode case in absence of the neoclassical closure (Fig. \ref{fig: NTM bootstrap current calculation}b), which confirms the meta-stability of the seed island in the large Lundquist number regime ($S\sim10^9$).

\begin{figure}[ht]
	\begin{center}
		\includegraphics[width=0.45\linewidth]{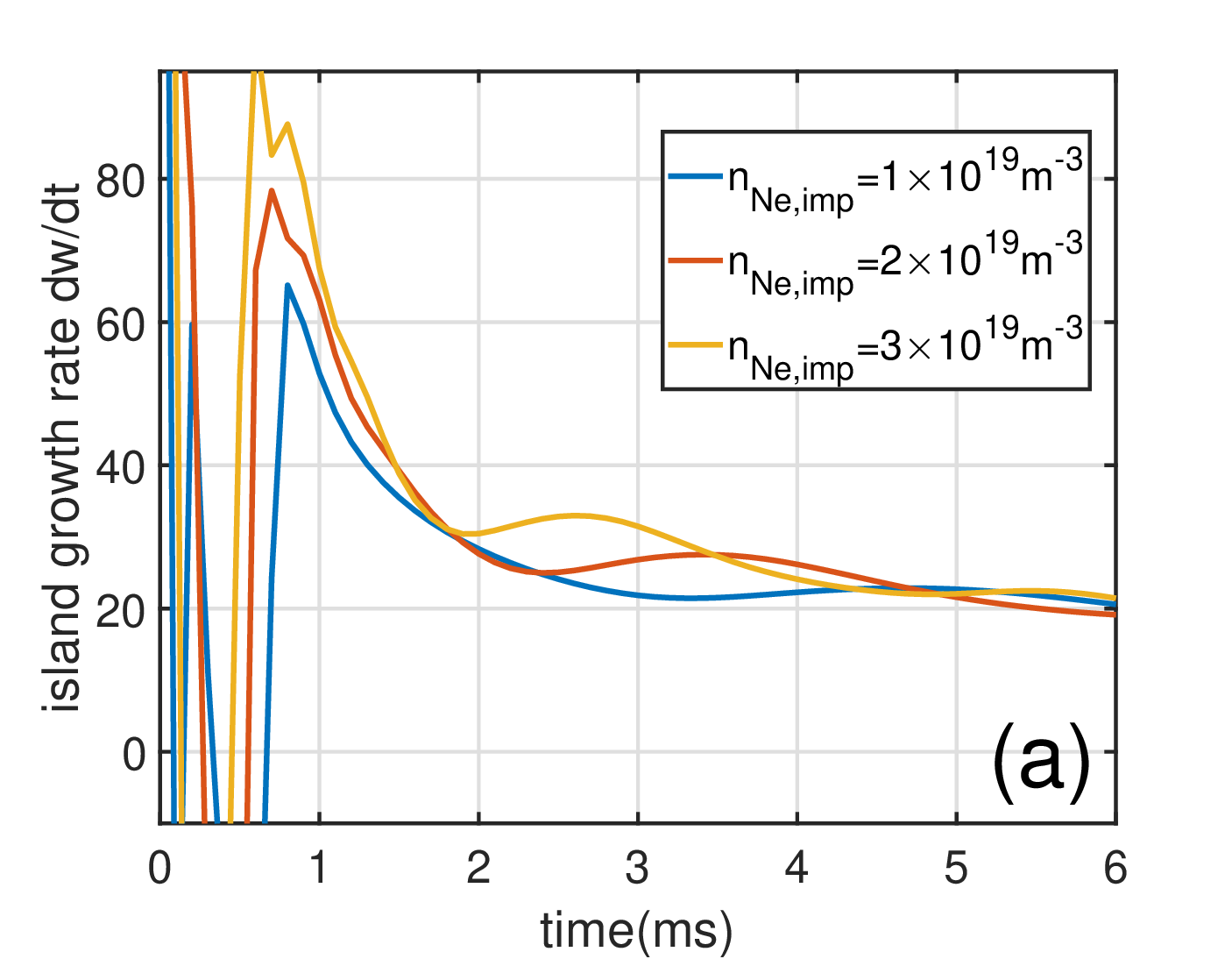}
		\includegraphics[width=0.45\linewidth]{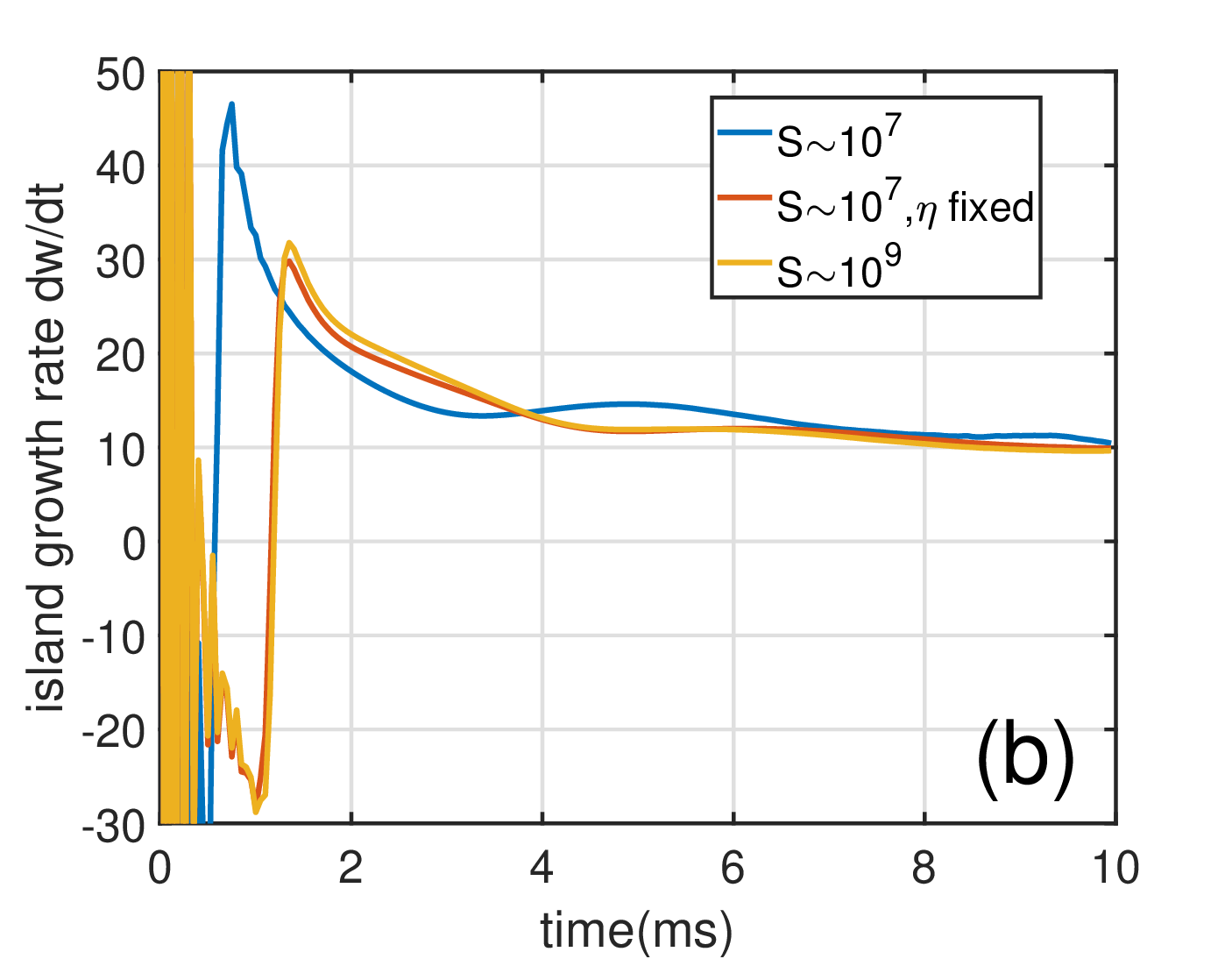}
	\end{center}
	\caption{The island growth rate of the $2/1$ mode as a function of time with various (a) impurity density levels, (b) Lundquist numbers and plasma resistivity models, where the coefficient $\mu_e=1\times10^7$s$^{-1}$.}
	\label{fig: NTM imp level}
\end{figure}

The simulation results with neoclassical closure presented in Fig. \ref{fig: NTM imp level}(a) show that higher impurity density level leads to faster island growth and larger island size, which confirms the expectation that faster nonlinear NTM growth can be driven by the larger radiation power with higher impurity density level, as a result of the larger perturbed bootstrap current $\delta J_{bs} \sim \sqrt{\epsilon}/B_{\theta}\left(\rm d\delta p/\rm dr\right)$ due to the enhanced pressure perturbation induced by stronger radiative cooling.
The island growth rates are nearly the same in various Lundquist number regimes and resistivity models, except for an earlier and faster growth in the low Lundquist number regime with Spitzer resistivity (Fig. \ref{fig: NTM imp level}b). The characteristic of the nonlinear tearing mode growth dominated by the neoclassical driving is less correlated to the plasma resistivity in comparison to the resistive tearing mode growth (Figs. \ref{fig: TM ne1e19 plasma resistivity}).



\section{Conclusion}
\label{Sec:conclusion}
In summary, this work demonstrates the seeding process through local impurity radiation cooling and the subsequent NTM onset using NIMROD simulation results.
The local helical current perturbation, which primarily comes from the diamagnetic current induced by the perturbed pressure gradient as a result of local radiative cooling, mainly drives the $\Delta^{'}$ from the outer region for the growth of seed island.
With a heuristic neoclassical closure included in the extended Ohm's law, the seed island is further driven by the perturbed bootstrap current from the electron neoclassical stress into the nonlinear island growth regime.

More sophisticated model, as well as the assessment of the influences due to NTM to the burning plasma performance, such as the value of $\beta_N$, alpha heating power, and the fusion power are planned in future work.
In particular, the coefficient $\mu_e$ in the heuristic electron stress tensor is set to be a constant in current model, which affects the intensity of the bootstrap current driving. The coefficient $\mu_e$ can be more accurately evaluated using \cite{Gianakon_2002}
\begin{equation}
	\mu_e\simeq-\frac{1.57f_T\nu_e}{\left( 1+1.07\nu_{*e}^{1/2}+1.02\nu_{*e}\right)\left( 1+1.07\nu_{*e}\epsilon^{3/2}\right)},
	\label{eq:mu_e formula}
\end{equation}
where $f_T$ is the trapped particle fraction, $\nu_{e}$ the electron collision frequency, $\nu_{*e}=\nu_{e}/\epsilon^{3/2}\omega_b$, and $\omega_b$ the bounce frequency. Particularly, the collision frequency, e.g. $\nu_{e}^{-1}\approx\left(12\pi^{3/2}/2^{1/2}\right) \left(m_e^{1/2}T_e^{3/2}\epsilon^2/n_iZ^2e^4\ln\Lambda\right)$, is strongly correlated to the charge number $Z$, the density and temperature perturbations, which should be greatly influenced by the impurity radiation cooling, hence critical to the bootstrap current driving. Although this work has demonstrated the main effects of the coefficient $\mu_e$, further update of the heuristic neoclassical closure is necessary for the more quantitative modeling of the NTM seeding and onset process.

\section*{Acknowledgments}
We are grateful for the supports from the NIMROD team and the CFETR physics design team. This work was supported by the National Magnetic Confinement Fusion Program of China (Grant No. 2019YFE03050004), the National Natural Science Foundation of China (Grant No. 51821005), and U.S. Department of Energy (Grant No. DE-FG02-86ER53218).
This research used resources of the National Energy Research Scientific Computing Center, a DOE Office of Science User Facility supported by the Office of Science of the U.S. Department of Energy under Contract No. DE-AC02-05CH11231 using NERSC award FES-ERCAP0027638.

\newpage
\section*{References}

\bibliographystyle{iopart-num}
\bibliography{ntm}

\newpage

\newpage

\newpage

\newpage

\newpage

\newpage

\newpage

%

\end{document}